\UseRawInputEncoding
\pdfoutput=1
\documentclass[letterpaper]{article}
\usepackage[letterpaper, margin=1in]{geometry}

\usepackage{amsmath}
\usepackage[dvipsnames]{xcolor}

\definecolor{red}{rgb}{0.8,0,0}
\definecolor{purered}{rgb}{1,0,0}
\definecolor{darkred}{rgb}{0.6,0,0}
\definecolor{green}{rgb}{0.0,0.5,0}
\definecolor{blue}{rgb}{0,0,0.75}
\definecolor{darkblue}{rgb}{0,0,0.55}
\definecolor{lightcyan}{rgb}{0.5,0.7,0.7}
\definecolor{orange}{rgb}{0.9,0.3,0.1}
\definecolor{purple}{rgb}{0.6,0.0,0.6}
\definecolor{cyan}{rgb}{0.0,0.7,0.7}
\definecolor{darkgray}{rgb}{0.4,0.4,0.4}
\definecolor{bronze}{rgb}{0.8, 0.5, 0.2}
\definecolor{dorange}{rgb}{0.75, 0.4, 0.0}
\definecolor{black}{rgb}{0.0,0.0,0.0}

\usepackage{multicol}
\usepackage{lipsum}
\usepackage{environ}
\usepackage{accents}
\usepackage{enumitem}

 \usepackage{subfiles} % Best loaded last in the preamble

\usepackage{array}
\usepackage{booktabs}
\usepackage[dvipsnames]{xcolor}
\usepackage{tikz}

\newcolumntype{M}[1]{>{\centering\arraybackslash}m{#1}}
 \usepackage{float}

 \usepackage[justification=centering]{caption}

\newfloat{video}{tbhp}{lst}%[section]
\floatname{video}{Video}
\def\Label{\refstepcounter{video}\label}

\usepackage[font={small},labelfont={bf}]{caption}
\usepackage{amssymb}

\definecolor{cvprblue}{rgb}{0.21,0.49,0.74}
\usepackage[pagebackref,breaklinks,colorlinks,citecolor=cvprblue]{hyperref}

\def\paperID{} % *** Enter the Paper ID here
\def\confName{}
\def\confYear{2024}

\title{Iterating the Transient Light Transport Matrix for Non-Line-of-Sight Imaging  \\[1em]

\includegraphics[width=1\linewidth]{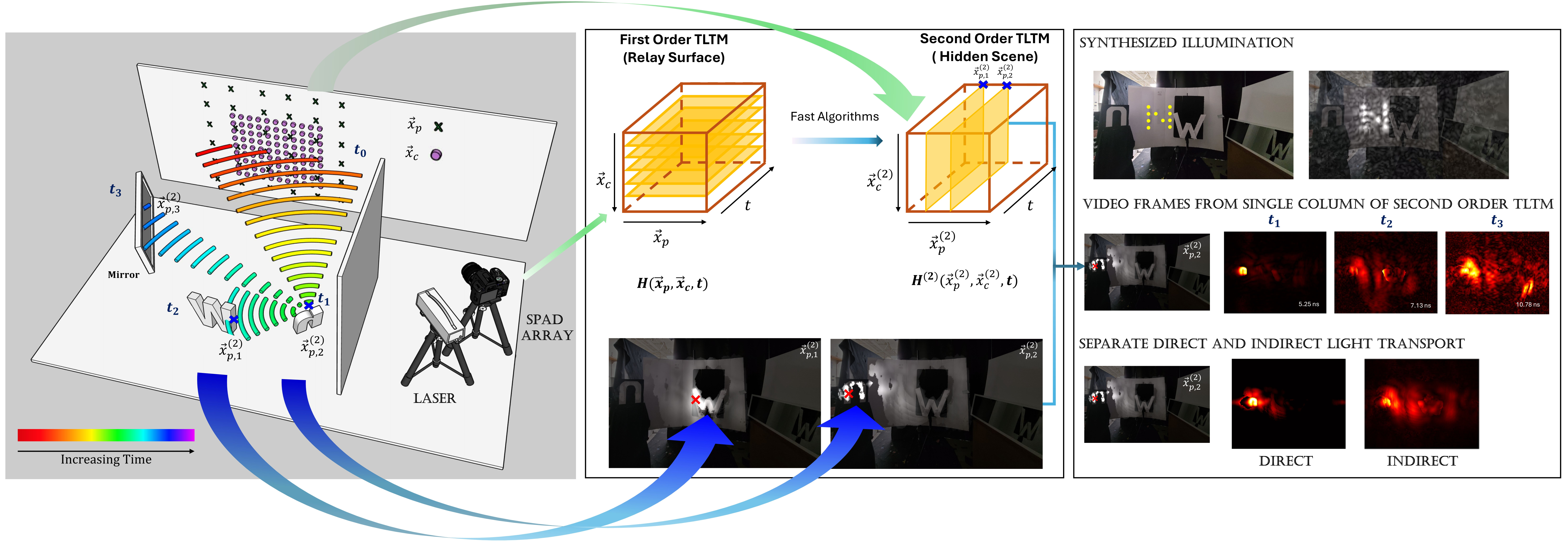}
\captionof{figure}{Emerging SPAD arrays enable the capture of the full Transient Light Transport Matrix (TLTM) of the Relay Surface in Non-Line-of-Imaging. Fast computational algorithms iterate on the TLTM and extract the second-order TLTM of the hidden scene around the corner. This is accomplished by computationally focusing our illumination on different parts of the hidden scene and then computing the corresponding spatiotemporal light transport. Here, we focus the illumination on the letter "n," which becomes illuminated at time $t_1$. Part of the light reflected from "n" then illuminates the left side of the letter "W" at time $t_2$, while the remaining light is reflected off a mirror at time $t_3$. The second-order TLTM can be used to extract complex light transport, relight the hidden scene under novel illumination, and separate the direct and indirect components of light transport around the corner.}
\label{fig:Teaser}}

   \author{Talha Sultan\\
University of Wisconsin - Madison\\
{\tt\small tsultan@wisc.edu}
\and
Eric Brandt \\
University of Wisconsin - Madison\\
{\tt\small elbrandt@wisc.edu}
\and
Khadijeh Masumnia-Bisheh\\
University of Wisconsin - Madison\\
{\tt\small khadijeh.masumnia-bisheh@wisc.edu}
\and
Simone Riccardo \\
Politecnico di Milano\\
{\tt\small simone.riccardo@polimi.it}
\and
Pavel Polynkin \\
University of Arizona\\
{\tt\small ppolynkin@optics.arizona.edu}
\and
Alberto Tosi \\
Politecnico di Milano\\
{\tt\small alberto.tosi@polimi.it}
\and
Andreas Velten \\
University of Wisconsin - Madison\\
{\tt\small velten@wisc.edu}
}

\begin{document}
\date{}
\maketitle
\begin{video}
\Label{vid:BM_FOCUS_SPATIAL}
\Label{vid:BM_FOCUS_SPATIAL_Raw}
\Label{vid:BM_FOCUS_Video_n}
\Label{vid:BM_FOCUS_Video_W}
\Label{vid:Complex_Light_Transport_Water}
\Label{vid:Complex_Light_Transport_Milk}
\end{video}

\setlength{\abovedisplayskip}{2pt}
\setlength{\belowdisplayskip}{2pt}

\clearpage

\begin{abstract}
Active imaging systems sample the Transient Light Transport Matrix (TLTM) for a scene by sequentially illuminating various positions in this scene using a controllable light source, and then measuring the resulting spatiotemporal light transport with time of flight (ToF) sensors. Time-resolved Non-line-of-sight (NLOS) imaging employs an active imaging system that measures part of the TLTM of an intermediary relay surface, and uses the indirect reflections of light encoded within this TLTM to “see around corners”. Such imaging systems have applications in diverse areas such as disaster response, remote surveillance, and autonomous navigation. While existing NLOS imaging systems usually measure a subset of the full TLTM, development of customized gated Single Photon Avalanche Diode (SPAD) arrays \cite{riccardo_fast-gated_2022} has made it feasible to probe the full measurement space. In this work, we demonstrate that the full TLTM on the relay surface can be processed with efficient algorithms to computationally focus and detect our illumination in different parts of the hidden scene, turning the relay surface into a second-order active imaging system. These algorithms  allow us to iterate on the measured, first-order TLTM, and extract a \textbf{second order TLTM for surfaces in the hidden scene}. We showcase three applications of TLTMs in NLOS imaging: (1) Scene Relighting with novel illumination, (2) Separation of direct and indirect components of light transport in the hidden scene, and (3) Dual Photography. Additionally, we empirically demonstrate that SPAD arrays enable parallel acquisition of photons, effectively mitigating long acquisition times.

\end{abstract}

% \section{Outline}
\iffalse
Outline
\begin{enumerate}
    \item Abstract
    \item Introduction
    \item Related Work?
    \item Methods
\end{enumerate}
\fi

\begin{figure*}[t]
  \centering
   \includegraphics[width=0.8\linewidth]{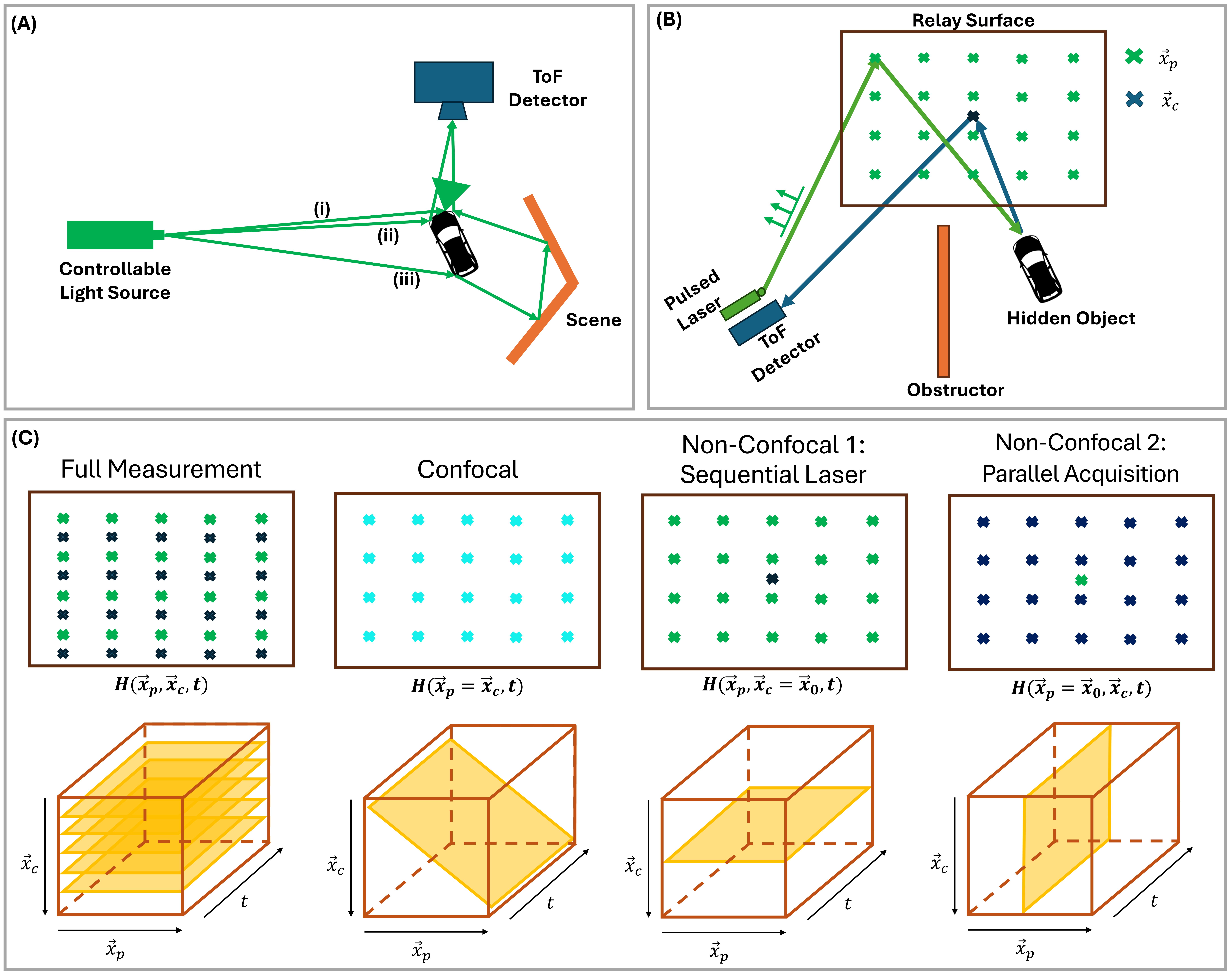}
   \caption{\textbf{(A)} Active Imaging System for measuring Transient Light Transport Matrix, which contains information about (i) Diffuse reflections (ii), specular reflections, and (iii) indirect components of Light Transport. \textbf{(B)} Non-line-of-sight Imaging System using Time of flight detectors is an active imaging system that measures the TLTM of the relay surface. \textbf{(C)} Existing acquisition schemes - Confocal and Non-Confocal - capture a subset of the Full Measurement. Multipixel arrays (Non-Confocal 2) can replicate data capture achieved by sequential laser scanning (Non-Confocal 1) while also reducing capture times through parallel acquisition of photons.}
   \label{fig:NLOS_Panel}
\end{figure*}

\section{Introduction}
\label{sec:intro}

Transient Light Transport Matrices (TLTM) characterize the spatiotemporal light transport from light sources to the detector for a given scene \cite{o2016optical}, and can be measured with an active imaging system (Fig. \ref{fig:Teaser}, Column 1) using a controllable light source and Time of Flight (ToF) detectors \cite{otoole_temporal_2014} (Fig. \ref{fig:NLOS_Panel}A). Once the TLTM is measured, different aspects of light transport can be manipulated in post-processing, such as scene relighting, separating direct from indirect light, specular from diffuse reflections, and even separating caustic from non-caustic effects \cite{paul1992synthetic, nayar_fast_2006, otoole_temporal_2014}.  

As the time resolution of ToF sensors improves \cite{velten_recovering_2012, buttafava_time-gated_2014}, it has become easier to extract higher-order information from the indirect components of light transport encoded within the TLTM. A key example of this is in Non-Line-of-Sight (NLOS) imaging systems \cite{velten_recovering_2012, faccio_non-line--sight_2020}, where the goal is to image objects hidden from the line of sight of an observer (Fig. \ref{fig:Teaser}, Column 1). One approach probes a subset of the 7D TLTM, denoted by $H(\vec{x_p}, \vec{x_c}, t)$, at an intermediary relay surface in the Line of Sight (LOS) of both the observer and the hidden object (Fig. \ref{fig:NLOS_Panel}B) by scanning the surface with a pulsed laser. At each illumination/laser position $\vec{x}_p$, photons scatter off the relay surface, interact with the hidden scene, and some photons scatter back to the relay surface where they are recorded at each detection/camera position $\vec{x}_c$ as a function of time. These "three bounce" photons from the hidden scene form the indirect component of the measured TLTM, and computational algorithms leverage this information to reconstruct the 3D geometry of the hidden scene \cite{liu_phasor_2020}. When the relay surface is planar and sampled with evenly spaced grids, the 7D TLTM collapses to a 5D TLTM, and computationally efficient Fourier-based algorithms can be used accelerate the reconstruction. 

Additional complex light transport, such as additional bounces or sub-surface scattering within the hidden scene, may be encoded in the TLTM, but is hard to extract with existing NLOS imaging systems. This is because these systems typically employ gated single-pixel Single Photon Avalanche Diode (SPAD) detectors \cite{buttafava_time-gated_2014, buttafava_non-line--sight_2015}, and collect a small subset of the full TLTM (Fig. \ref{fig:NLOS_Panel}C). While this information is sufficient to generate 3D reconstructions, this limited measurement makes it difficult to extract rich information pertaining to the light transport available within the optical signal at the relay surface. 

With the advent of new gated SPAD arrays \cite{renna_fast-gated_2020, riccardo_fast-gated_2022, zhao_gradient-gated_2023}, it has become feasible to fully probe this TLTM, opening exciting avenues of research. In this work, we build a NLOS system with a 16 x 16 gated SPAD array \cite{riccardo_fast-gated_2022} that is able to sample all dimensions of the full 7D TLTM (Fig. \ref{fig:NLOS_Panel}C, Column 1). Furthermore, we draw inspiration from phased array systems to develop fast algorithms that computationally focus and image our virtual illumination at 3D locations, ${\vec{x}_p}^{(2)}$ and ${\vec{x}_c}^{(2)}$ respectively, in the hidden scene. This focusing ability allows us to iterate on the first order TLTM, $H(\vec{x_p}, \vec{x_c}, t)$, and extract a second order transient light transport matrix, $H^{(2)}\left(\vec{x}_p^{(2)}, \vec{x}_c^{(2)}, t\right)$, \textbf{\textit{for surfaces in the hidden scene}} (Fig.\ref{fig:Teaser}, Column 2). This provides a formalism for reconstructing scenes around multiple corners by iteratively reconstructing and using higher order TLTMs
$H^{(n)}\left(\vec{x}_p^{(n)}, \vec{x}_c^{(n)}, t \right)$.

\paragraph{Contributions} We can summarize our contributions as follows: 
\begin{itemize}
    \item We build a NLOS imaging system with a gated  16x16 SPAD array \cite{riccardo_fast-gated_2022} that is able to capture the full first order TLTM, i.e. TLTM-1.
    \item We develop fast, computationally efficient algorithms that can iterate on TLTM-1 and extract the second order TLTM, i.e. TLTM-2, for the hidden scene. We demonstrate that complex light transport, such as subsurface scattering and shadows, are preserved in the TLTM-2 
    \item We demonstrate different applications with this higher order TLTM, such as separation of direct and indirect light transport, scene relighting with novel illumination, and dual photography (Fig.\ref{fig:Teaser}, Column 3).
\end{itemize}

\section{Related Work}
\label{sec:RelatedWork}
\paragraph{NLOS Imaging with ToF Detectors} While many different approaches have emerged to tackle the problem of NLOS imaging \cite{maeda_recent_2019,  faccio_non-line--sight_2020, geng_recent_2022}, NLOS imaging using photon time of flight has demonstrated 3D reconstructions of a wide variety of complex room-sized scenes \cite{lindell_wave-based_2019, liu_non-line--sight_2019} and relay surfaces \cite{lindell_wave-based_2019, manna_non-line--sight-imaging_2020, liu_non-line--sight_2023} with high capture speeds \cite{pei_dynamic_2021, nam_low-latency_2021}, and even reconstruct objects hidden around two corners \cite{royo_virtual_2023}. 

\paragraph{Phasor Field Framework} The development \cite{reza_phasor_2019-1, teichman_phasor_2019, dove_paraxial_2019, dove_paraxial_2020, dove_nonparaxial_2020, dove_speckled_2020, dove2020theory} and validation \cite{reza_phasor_2019, reza_phasor_2019-1, sultan_towards_2024} of the phasor field framework has enabled the modeling of an NLOS imaging system as a virtual LOS imaging system at the relay surface. In this framework, light transmitted from and received at the relay surface is described as a virtual wavefront. This allows us to use familiar intuition and methodology from other LOS imaging sytems such as cameras, microwave phased arrays, or ultrasound imaging systems to understand the problem. Diffraction integrals from classical wave optics correctly describe the propagation of the virtual wavefront, and can be leveraged to backpropagate it and generate 3D reconstructions of the hidden scene \cite{liu_non-line--sight_2019}. It also correctly models the propagation of light inside the scene and explains phenomena like the NLOS missing cones problem \cite{liu_analysis_2019, royo_virtual_2023} and the formation of virtual mirror images in NLOS reconstructions \cite{royo_virtual_2023}. 

\paragraph{Hardware for ToF NLOS} One approach demonstrating ToF NLOS imaging utilized a femtosecond pulsed laser and a Streak camera \cite{velten_recovering_2012} to generate 3D reconstructions of the hidden scene. Subsequent work developed low-cost NLOS imaging systems using modulated laser diodes for illumination and photonic mixer devices (PMDs) for detection \cite{heide_low-budget_2013, heide_diffuse_2014}. To reduce hardware costs without substantially sacrificing time resolution, a picosecond pulsed laser was later paired with a customized single-pixel SPAD sensor for NLOS imaging \cite{buttafava_non-line--sight_2015}. This SPAD sensor was adapted for NLOS imaging by incorporating a time-gating feature that selectively captures relevant third-bounce photons from the hidden scene while gating out first-bounce photons from the relay surface. This minimizes distortion in the third-bounce signal due to photon pile-up from direct reflections \cite{becker_advanced_2005}.

These single-pixel SPAD systems usually acquire a subset of the TLTM of the relay surface using one of two acquisition schemes. Confocal acquisition scheme involves scanning the relay wall with both the laser and the single-pixel detector focused on the same spot on the relay surface, i.e., $H(\vec{x}_p = \vec{x}_c, t)$, capturing only the diagonal of TLTM-1 (Fig. \ref{fig:NLOS_Panel}C, Column 2). In non-confocal acquisition, the single pixel detector remains at a fixed relay wall location $\vec{x}_0$ while the laser scans the relay surface independently $H(\vec{x}_p, \vec{x}_c = \vec{x}_0, t)$, measuring a row of TLTM-1 (Fig. \ref{fig:NLOS_Panel}C, Column 3). During reconstruction, Helmholtz reciprocity \cite{veach_robust_1997, sen_dual_2005} is often used to map this row to a column by applying a transpose operation: i.e. $H(\vec{x}_p, \vec{x}_c = \vec{x}_0, t) = H^T(\vec{x}_c = \vec{x}_0, \vec{x}_p , t)$. This operation interchanges the illumination and detection positions for the reconstruction operator so the SPAD pixel acts as a virtual point light source and the laser positions on the relay surface, $\vec{x}_c$, form a virtual aperture. Both acquisition schemes capture a subset of the full measurement space (Fig. \ref{fig:NLOS_Panel}C), enabling 3D reconstructions of the hidden scene geometry but restricting the system's ability to capture and extract higher-order light transport. Another limitation of these systems is that the optical signal from the hidden scene is very weak and the relay surface must be scanned for long durations or with the laser power increased to unsafe levels. 

\paragraph{SPAD arrays for ToF NLOS} 
Multi-pixel SPAD arrays are emerging as a viable technology \cite{renna_fast-gated_2020,riccardo_fast-gated_2022, zhao_gradient-gated_2023} that enable efficient capture of TLTM-1 through parallel acquisition of photons at different parts of the relay surface (Fig. \ref{fig:NLOS_Panel}C), reducing the need for long exposures or high laser power. However, recent work utilizing SPAD arrays typically focuses on capturing a column of TLTM-1 either directly \cite{pei_dynamic_2021}, through remapping \cite{nam_low-latency_2021}, or using Helmholtz reciprocity \cite{royo_virtual_2023} without sampling all available dimensions of the TLTM-1. 

For example, a 32 x 32 pixel SPAD array was paired with a laser focused at a single illumination position \cite{pei_dynamic_2021} to capture a column of the TLTM-1 directly, enabling reconstructions of small hidden objects at 20 Hz. Another approach paired two 16 x 1 pixel SPAD arrays \cite{renna_fast-gated_2020} with a sparse laser grid to achieve live NLOS video reconstructions of room-sized hidden scenes at 5 Hz \cite{nam_low-latency_2021}. Their approach relies on remapping the measurement to a single column of the TLTM. Other work \cite{gu_fast_2023, royo_virtual_2023} used a 16 x 16 SPAD array \cite{riccardo_fast-gated_2022} but only measured a row of the TLTM with high SNR by focusing the array on a small area (7.1 cm x 4.7 cm) of the relay surface. While this set up is able to generate reconstructions of objects hidden around two corners using 5th bounce photons \cite{royo_virtual_2023} with reasonable acquisition times, their framework relies on having flat surfaces in the hidden scene that act as "virtual mirrors" to reconstruct the mirror \textit{image} of hidden object without computing TLTM-2 of the hidden scene. To simultaneously detect both the first bounce from the relay surface for calibration and the 3rd bounce from the hidden scene while mitigating pile-up, a recent approach applies a variable gate to individual SPAD pixels of a 6 x 6 SPAD array \cite{zhao_gradient-gated_2023}. All these systems capture a lower-dimensional subset of the full TLTM (Fig. \ref{fig:NLOS_Panel}C).

\paragraph{Phased Array Systems and Beamforming} After capturing the full TLTM-1, each point in the illumination grid focused on the relay wall can be modeled as an omnidirectional element within a transmitting phased array, sending phasor field waves in all directions into the hidden scene. Simultaneously, each point in the detection grid functions as an omnidirectional element within a separate receiving phased array, capturing the waves reflected from the hidden scene. By applying beamforming principles, we can synthesize illumination using the transmitting array and reconstruct an image of the hidden scene based on data captured by the receiving array. This approach transforms the relay surface into an active, coherent imaging system that illuminates and images the hidden scene from the perspective of the relay wall. This framework allows us to draw parallels between NLOS imaging systems and phased array systems. For example, the Total Focusing Method (TFM) employed in ultrasound phased array systems \cite{holmes_post-processing_2004} is mathematically equivalent to the backprojection algorithm \cite{velten_recovering_2012}. Additionally, measuring the full TLTM-1 is equivalent to MIMO systems in RADAR \cite{li_mimo_2008} or full matrix capture in ultrasound phased arrays \cite{zhang_defect_2010}. In this work, we sample all dimensions of both the illumination and the detection grids, measuring the full TLTM-1. This enables us to synthetically focus the phasor field waves in the hidden scene \cite{schmerr_fundamentals_2015} using a signal processing framework \cite{van_veen_beamforming_1988}, and extract TLTM-2 of surfaces in the hidden scene. 

\paragraph{Light Transport Matrices} The Light Transport Matrix (LTM) characterizes the spatial light transport from illuminating any pixel in a given scene to all the other pixels. It has extensively been used in computer graphics where a controllable light source, typically a projector, and a detector, typically a camera, are used to measure the LTM for a given scene (Fig. \ref{fig:NLOS_Panel}A). Once the LTM is measured, the image can be manipulated during post-processing. For instance, dual photography leverages the transpose of the LTM to swap the roles of illumination and detection, enabling the scene to be imaged from the perspective of the light source \cite{sen_dual_2005}. Additionally, the LTM is useful for separating direct and indirect components of light transport \cite{nayar_fast_2006, otoole_primal-dual_2012} and synthetic scene relighting. \cite{paul1992synthetic, o2016optical}.

Transient Light Transport Matrix (TLTM) extend LTMs by tracking both spatial and \textit{temporal} evolution of light transport, and have become possible to capture as the timing accuracy of ToF sensors has improved. \cite{otoole_temporal_2014}. The inclusion of the time dimension allows for easier separation of direct and indirect components of light, while the temporal frequency domain of the TLTM can be used to distinguish between caustic and non-caustic effects \cite{otoole_temporal_2014}. 

\paragraph{TLTMs in ToF NLOS Imaging} ToF NLOS imaging systems typically capture a row or diagonal of TLTM of the relay surface sequentially, and process the indirect component to generate 3D reconstructions of the hidden scene. This sequential, basis scan is the slow way to probe the first order TLTM \cite{sen_dual_2005, wang_kernel_2009, otoole_optical_2010}, but is feasible for NLOS imaging since the number of illumination points is 4 orders of magnitude smaller than in megapixel projectors commonly used for LOS imaging systems. 

Marco et al \cite{marco_virtual_2021} use the phasor field framework to build Virtual Light Transport Matrices (VLTM) that can be used to separate the direct and indirect components of light transport in the hidden scene. However, their method does not extract the entire \textit{Transient} Light Transport Matrix since VLTMs do not capture the temporal evolution of light transport in the hidden scene. Additionally, they employ algorithms that are computationally intensive, making it infeasible to extract and differentiate additional bounces within the indirect component of the light transport that occurs within the hidden scene. Their experimental hardware scans a sparse laser grid and detects light with linear SPAD arrays, preventing the sampling of all dimensions of TLTM-1. In this work, we develop algorithms that generate the complete Transient Light Transport Matrix with lower computational complexity (see Section \ref{sec:Complexity}) by utilizing the Fast Fourier Transforms (FFT) to speed up the computation. Additionally, we build an NLOS set up that employs a 2D SPAD array \cite{riccardo_fast-gated_2022} to efficiently sample all dimensions of TLTM-1, and use this dataset to compute the complete TLTM-2.

\section{Methods - TLTM Generation}
In this section, we show that the second order TLTM can be computed from the measured TLTM on the relay surface using a series of linear operators denoted by $f$:

\begin{equation}
% \label{eq:FBP_PF_Omega_1}
    \begin{split}    H^{(2)}\left(\vec{x}_p^{(2)}, \vec{x}_c^{(2)}, t \right)  &= f(H\left(\vec{x}_p, \vec{x}_c, t \right)) \\
    \end{split}   
\end{equation}

Generating TLTM-2 requires that we sequentially focus our illumination to a given point, $\vec{x}_p^{(2)}$, in the hidden scene, and image the hidden scene at voxels $\vec{x}_c^{(2)}$ as a function of time. To focus our illumination, we utilize the phasor field framework which converts the relay surface into a virtual active LOS imaging system that illuminates the hidden scene from the relay surface with arbitrary, virtual illumination described by a projector function, $\mathcal{P}\left(\vec{x}_p, t\right)$  and images the hidden scene using some camera function, $\Phi(.)$, that describes the imaging operation (see Fig S.1 in \cite{liu_non-line--sight_2019}):

\begin{equation}
I\left(\vec{x}_v, t \right)= \Phi \left( \left[\mathcal{P}\left(\vec{x}_p, t \right) * H\left(\vec{x}_p, \vec{x}_c, t\right)\right] \right).
\label{eq:PF_Camera}
\end{equation}

where $\vec{x}_v$ is a voxel that defines a 3D location in the hidden scene. To extract TLTM-2, we set the projector function such that it creates a creates a focusing wave in space-time at $\vec{x}_p^{(2)}$ (details in supplement section \ref{sec:PF_Derivations}), and then the following camera function can be used to extract the a single column of TLTM-2 by operating on the temporal frequency ($\omega$) domain of the virtual response:

\begin{equation}
\begin{aligned}  
H^{(2)}(\vec{x}_p^{(2)}, \vec{x}_c^{(2)}, t) &= \iiint   \mathcal{P}(\omega) H\left(\vec{x}_p, \vec{x}_c, \omega\right) e^{-j \omega \Delta t  \left(\vec{x}_p^{(2)}\right)}  \\ &   e^{-j \frac{\omega}{c}\left|\vec{x}_c^{(2)}-\vec{x}_c \right| }  e^{j\omega t} d \vec{x}_p d\vec{x}_c d\omega \\ 
&= f(H\left(\vec{x}_p, \vec{x}_c, t \right))
\label{eq:BM_Video}
\end{aligned}
\end{equation}

where $\Delta t \left(\vec{x}_p^{(2)}\right)$ adds a spherical curvature to the temporal wavefront (details in Supplement Section \ref{sec:BM_Derivations}) that focuses the virtual illumination, and the linearity of the integral operator implies that $f$, which extracts TLTM-2 from TLTM-1, is itself a linear operator. The computation can be accelerated by first incorporating focusing timeshifts and integrating over $\vec{x}_p$ to generate $\mathcal{P_F} \left(\vec{x}_c, \omega\right)$:

\begin{equation}
\begin{aligned}  
H^{(2)}( \vec{x}_p^{(2)},  & \vec{x}_c^{(2)}, \ t)
   \\  &= \iint e^{j\omega t} e^{-j \frac{\omega}{c}\left|\vec{x}_c^{(2)}-\vec{x}_{\mathrm{c}}\right|}
     \\ &  \underbrace{ \left[\int \mathcal{P}(\omega) H \left(\vec{x}_p, \vec{x}_c, \omega\right)
   e^{-j \omega \Delta t \left(\vec{x}_p^{(2)}\right)}
   d \vec{x}_{\mathrm{p}} 
    \right]}_{\mathcal{P_F} \left(\vec{x}_c, \omega\right)} d \omega  d \vec{x}_{\mathrm{c}}   \\
    &= \int e^{j\omega t} \underbrace{ \int \left[ \mathcal{P_F} \left(\vec{x}_c, \omega\right) e^{-j \frac{\omega}{c}\left|\vec{x}_c^{(2)}-\vec{x}_{\mathrm{c}}\right|} d \vec{x}_{\mathrm{c}}  
    \right]}_{R_{x_v} (\mathcal{P_F} \left(\vec{x}_c, \omega\right)) } d \omega  
\label{eq:BM_Video_Fast}
\end{aligned}
\end{equation}

then Fourier based convolution can be used to accelerate $R_{x_v} (\mathcal{H_F} \left(\vec{x}_c, \omega \right))$, provided $\vec{x}_c$ is sampled on a regular grid (see Supplement Section \ref{sec:PF_Derivations}). The overall complexity for running Eq \ref{eq:BM_Video} once to generate a single column of TLTM-2 is $O(kN^3log(N))$, where $k$ is the number of transient frames in our 4D video, and $N$ is the side length of the reconstructed volume \cite{liu_phasor_2020} (see Supplement Section \ref{sec:Complexity}). 

\subsection{Beamforming Resolution}
\label{sec:BM_Res}
Under the phasor field framework, both the focusing and imaging resolution can shown to be governed by Rayleigh Criterion \cite{goodman_introduction_2005, hecht_optics_2017, dove_paraxial_2019, dove_paraxial_2020, sultan_towards_2024}: 

\begin{equation}
\begin{aligned}  
\Delta x &= \frac{1.22\lambda_c z}{D} 
\label{eq:Focusing_Res}
\end{aligned}
\end{equation}

where $\lambda_c = c/2\pi \omega_0$ is the central the phasor field wavelength (see Section \ref{sec:PField_Projector_Function}), $z$ is the depth offset from the relay wall, and $D$ is the size of either the illumination or detector grid to yield focusing or imaging resolution respectively.

Beamforming using phased array principles requires sampling the relay surface at the Nyquist rate, given by $\lambda_s = \lambda_c/2$. For a given number of grid positions $N$, the spatial size of the grid on the wall can be rewritten as $D = N\cdot \lambda_s$. Substituting this into Eq \ref{eq:Focusing_Res} and simplifying:

\begin{equation}
\begin{aligned}  
\Delta x &= \frac{1.22 z}{2N} 
\label{eq:Focusing_Res_SPAD}
\end{aligned}
\end{equation}

so the resolution ultimately only on the number of grid positions $N$ and depth $z$. Since the illumination positions are an order of magnitude larger than the number of detector positions for our setup, the resolution for this system is ultimately limited by the number of SPAD pixels. 

Beamforming can also gives rise to unfavorable side lobes \cite{schmerr_fundamentals_2015}, and together with an insufficient amount of SPAD pixels may limit focusing resolution, which then determines how well we can probe the second order TLTM. 

In the next two subsections, we utilize two straightforward techniques to improve focusing resolution to maximize the quality of the extracted TLTM-2. 

\subsubsection{Grid Interpolation}
To maximize the extraction of second order TLTM with limited SPAD pixels, we interpolate $ H\left(\vec{x}_p, \vec{x}_c, \omega\right)$ along $\vec{x}_p$ and reconstruct with a smaller wavelength $\lambda_c$. As indicated by Eq \ref{eq:Focusing_Res_SPAD}, this makes the focus sharper, and works well in practice since the phasor field wavefront varies slowly in the transverse dimension along the relay surface \cite{gu_fast_2023}.

\subsubsection{Side Lobe Suppression}
Beamforming artifacts, such as unwanted side lobes or residual illumination (see Section \ref{sec:TLTM_Extract}), arise due to the irregularities in the SPAD grid from non uniform spacing or holes due to presence of non-functional pixels. An alternate way to improve the quality of our illumination in the hidden scene is to solve an additional linear inverse that adjusts the projector function to suppress these artifacts in the focused illumination. While the direct focusing equation, Eq \ref{eq:curvature_parabolic} imparts a spherical curvature to the wavefront to focus to given point and works well for uniform phased arrays, the linear inverse infers the optimal wavefront shape at the relay surface that best focuses to a desired location, $\vec{x}_p^{(2)}$, in the hidden scene, given the spatial location and sensitivity of each SPAD pixel.

The algorithm proceeds as follows. First, we compute the time, $t_i = |\vec{x}_p^{(2)} - \bar{x}_p|/c$, at which the virtual illumination from the mean SPAD array position, $\bar{x}_p$, reaches the desired focus point, $\vec{x}_p^{(2)}$. Then, we sequentially compute and save the 4D video at time $t_i$ separately for each SPAD pixel using an adapted form of Eq \ref{eq:Video_Voxel_Freq_Offset}:

\begin{equation}
\begin{split} 
{\mathcal{A}}\left(\vec{x}_v, \vec{x}_p, t' = t_i \right)  &= \iint 
    \left[\mathcal{P}(\omega)  H\left(\vec{x}_p,\vec{x}_c, \omega\right) \right] \\
      & e^{-j\omega t_i} e^{-j \frac{\omega}{c}\left(\left|\vec{x}_c-\vec{x}_v\right|\right)} d \omega d \vec{x}_c  \\
\end{split}  
\label{eq:A_Matrix_Space}
\end{equation}

where $\mathcal{P}(\omega)$ is the unmodified projector function defined in Eq \ref{eq:PFPacket_Omega}, $\vec{x}_v$ is a point in the hidden scene, and $\mathcal{A}$ is a 2D matrix with the $i^{\text{th}}$ column being the reconstruction at $t = t_i$ for a given illumination pixel position, $\vec{x}_{p_j}$ on the relay surface. Let $\mathcal{B}$ be a pattern such that 

\begin{equation}
\begin{split} 
\mathcal{B}(\vec{x}_v) =
\begin{cases}
1, & \vec{x}_v = \vec{x}_p^{(2)} \\
0, & \textit{otherwise} \\
\end{cases}
\end{split}  
\label{eq:B_Matrix_Space}
\end{equation}

Now, we can compute the complex coefficients $\mathcal{X}$:

\begin{equation}
\begin{aligned}  
\mathcal{X}   = \left(\mathcal{A}^{T} \mathcal{A} + \lambda_R I\right )^{-1} \mathcal{A} \mathcal{B}
\label{eq:BM_Pattern_Space}
\end{aligned}
\end{equation}

where $\lambda_R$ prevents overfitting to the noise, and the $i^{\text{th}}$ entry of $\mathcal{X}$ is a complex coefficient, $\alpha \left(\vec{x}_p,\vec{x}_p^{(2)}\right)$, that adjusts the complex wavefront of the $i^{\text{th}}$ SPAD pixel spatially to generate the best possible focus at $\vec{x}_p^{(2)}$. This can be done by incorporating the following optimized projector function into Eq \ref{eq:Video_Voxel_Freq_Offset}: 

\begin{equation}
\begin{aligned} 
\mathcal{P}\left(\vec{x}_p, \omega \right) =  \alpha \left(\vec{x}_p,\vec{x}_p^{(2)}\right) \mathcal{P}\left(\omega\right)
\label{eq:Projector_Space}
\end{aligned}
\end{equation}

\section{Applications of Second Order TLTM}
After computing the second-order TLTM, we can extract various aspects of light transport to showcase different applications, which we outline in this section.

\subsection{Scene Relighting with Novel Illumination}
We can adjust our projector function to isolate the reconstruction from a single SPAD pixel, $\vec{x}_p = {\vec{x}_0}$:

\begin{equation}
\begin{aligned} 
\mathcal{P}\left(\vec{x}_p, \omega \right) &=  \delta (\vec{x}_p - \vec{x}_0) \mathcal{P}\left(\vec{x}_p, \omega \right) \\
&= \mathcal{P}\left(\vec{x}_0, \omega \right)
\end{aligned}
\end{equation}

Since each SPAD pixel illuminates the hidden scene from a unique location, we can incorporate this modified projector function into Eq. \ref{eq:Voxel_Freq} to save 3D reconstructions, $ I\left( \vec{x}_v, \vec{x}_p \right)$, of the hidden scene under novel illumination angles.  Additionally, by incorporating this modified projector function into Eq. \ref{eq:Video_Voxel_Freq}, we can create 4D videos, $I\left( \vec{x}_v, \vec{x}_p, t \right)$, showing our SPAD pixels illuminating the hidden scene from different angles over time. 

Adding up the reconstructions incoherently, by summing up the individual magnitudes, simulates the result of illuminating the hidden scene using an array of incoherent point light sources on the relay surface.  In contrast, adding the reconstructions coherently enables the design of optimal projector functions that convert points on hidden surfaces into virtual point light sources. Then, we can synthesize arbitrary illumination patterns as a combination of these individual point sources.

\subsubsection{Relighting with Synthesized Illumination}
We can create a coherent point light source in the hidden scene by synthetically focusing the illumination using coherent beamforming outlined in Eq \ref{eq:BM_Video}. Alternately, we can optimize our projector function over $\vec{x}_p$ repeatedly by recomputing the complex coefficients, $\mathcal{X}$, in Eq \ref{eq:BM_Pattern_Space} to focus on individual points in the hidden scene. Retaining temporal coherence for the point light source in the hidden scene is important for iterating on TLTM-2 and extracting higher order TLTM. In either case, the output can be combined linearly to generate arbitrary spatial patterns. However, the spatial resolution is limited by the number of SPAD pixels (Eq \ref{eq:Focusing_Res_SPAD}), and is too low for our current set up to generate meaningful patterns.

\subsubsection{Scene Relighting with Incoherent Point Source}
To gain further control over the synthesized illumination, we can improve spatial resolution by sacrificing temporal coherence. This is accomplished by optimizing the projector function over both $\vec{x}_p$ and $\omega$, to focus to a point, $\vec{x}_v$ in the hidden scene. We modify equation Eq \ref{eq:A_Matrix_Space} to generate a matrix $\mathcal{A}_{*}$: 

\begin{equation}
\begin{split} 
{\mathcal{A}_{*}}\left(\vec{x}_v, \vec{x}_p, \omega, t' = t_i \right)  &= \int 
    \left[\mathcal{P}(\omega)  H\left(\vec{x}_p,\vec{x}_c, \omega\right) \right] \\
      & e^{-j\omega t_i} e^{-j \frac{\omega}{c}\left(\left|\vec{x}_c-\vec{x}_v\right|\right)} d \vec{x}_c  \\
\end{split}  
\label{eq:A_Matrix_Space_Time}
\end{equation}

and we reshape $\mathcal{A}_{*}$ such that the $i^{\text{th}}$ column of $\mathcal{A}_{*}$ is the reconstructed surface for a given illumination position ${\vec{x}_p}$, and frequency $\omega$. Let $\mathcal{B}$ be a pattern such that 
 
\begin{equation}
\begin{split} 
\mathcal{B}(\vec{x}_v) =
\begin{cases}
1, & \vec{x}_v = {\vec{x}_p}^{(2)} \\
0, & \textit{otherwise} \\
\end{cases}
\end{split}  
\label{eq:B_PATTERN}
\end{equation}

Now, we can compute the complex coefficients $\mathcal{X}_{*}$:

\begin{equation}
\begin{aligned}  
\mathcal{X}_{*}   = \left({\mathcal{A}^{T}_{*}} \mathcal{A}_{*} + \lambda_R I\right )^{-1} \mathcal{A}_{*} \mathcal{B}
\label{eq:BM_Pattern_Space_Time}
\end{aligned}
\end{equation}

where the $i^{\text{th}}$ entry of $\mathcal{X}_{*}$ is a complex coefficient, $\alpha\left(\vec{x}_p, \omega, \vec{x}_p^{(2)}\right)$, corresponding to the $i^{\text{th}}$ column of $\mathcal{A}_{*}$. We can define a new projector function: 

\begin{equation}
\begin{aligned} 
\mathcal{P}(\vec{x}_p, \omega) = \alpha \left(\vec{x}_p, \omega, \vec{x}_p^{(2)}\right) \mathcal{P}(\omega)
\label{eq:Projector_Space_Time}
\end{aligned}
\end{equation}

and generate the desired pattern on the surface by incorporating this modified projector function into Eq \ref{eq:Video_Voxel_Freq_Offset}: 

 \begin{equation}
\label{eq:PF_RSD_SS_Patterns}
    \begin{split}  I_\mathcal{P} \left(\vec{x}_v, t^{\prime} = t_i\right)  &= \iiint  \left[ \alpha \left(\vec{x}_p, \omega, \vec{x}_p^{(2)}\right) \mathcal{P}(\omega)  H\left(\vec{x}_p, \vec{x}_c, \omega\right) \right] \\
    & e^{-j\omega t_i} e^{-j \frac{\omega}{c}\left(\left|\vec{x}_c-\vec{x}_v\right|\right)}d \vec{x}_p d \vec{x}_c d\omega.     \\
    &= \phi \left(\vec{x}_v, \vec{x}_p^{(2)}\right)
    \end{split}  
\end{equation}

This generates an incoherent point light source at $\vec{x}_v = \vec{x}_p^{(2)}$ with spatial resolution that is an order of magnitude better that of a coherent point light source. To synthesize arbitrary patterns, we decompose the pattern into multiple $\vec{x}_{p_i}^{(2)}$ and repeat the computation for Eq \ref{eq:A_Matrix_Space_Time} $-$ \ref{eq:PF_RSD_SS_Patterns}. The matrix $A_{*}$ is large, and must be regenerated if $t_i$ varies across $\vec{x}_p^{(2)}$, making this process computationally expensive  (see Section \ref{sec:Complexity_Projector}). The final pattern, $I_\mathcal{P} \left(\vec{x}_v\right)$, can be computed using:

\begin{equation}
\label{eq:Pattern_Normalize}
I_\mathcal{P} \left(\vec{x}_v\right)  = \sum_i \eta_i \phi\left(\vec{x}_v, \vec{x}_{p_i}^{(2)}\right)
\end{equation}

where $\eta_i = \left| \phi\left(\vec{x}_v = \vec{x}_{p_i}^{(2)}, \vec{x}_{p_i}^{(2)}\right) \right|^{-1}$ is a normalization constant that improves visualization.

While we can use the entire volume, voxels with intensities close to zero are not useful since multiplying any complex number, $\alpha(\vec{x}_p, \omega)$, by zero will still yield zero. Therefore, we find and save voxels, $\vec{x}_s$, corresponding to a 2D surface by applying a max filter along depth. Then, we generate our pattern on these voxels using Eq \ref{eq:A_Matrix_Space_Time} $-$ \ref{eq:PF_RSD_SS_Patterns}.

\subsection{Separation of Direct and Indirect Components }

With the TLTM, the direct and indirect components can be separated efficiently with computational Time Gating on the temporal dimension. For each voxel, $x_v$, we can compute the time of flight, $t_i = |x_v - \bar{x}_p|/c + K$ , for the illumination to reach $x_v$ from the center of the SPAD array $\bar{x}_p$, i.e, the center of the illumination pulse. We add a constant $K = \sigma$ from Eq \ref{eq:PFPacket_time}, to account for the temporal width of the illumination pulse as it propagates in time. Then, the direct component is the sum of video frames that occur before this time $t_i$ and the indirect component is the sum of video frames that occur after $t_i$ i.e. 

\begin{equation}
\begin{aligned}  
 I_\mathcal{D} \left(\vec{x}_{v_p},
    \vec{x}_{v_c},t \right) &= \int M_1 (t)    I_\mathcal{P} \left(\vec{x}_{v_p},
    \vec{x}_{v_c}, t \right) dt 
\label{eq:BM_Direct}
\end{aligned}
\end{equation}

\begin{equation}
\begin{aligned}  
 I_\mathcal{I} \left(\vec{x}_{v_p},
    \vec{x}_{v_c},t \right) &= \int M_2 (t)    I_\mathcal{P} \left(\vec{x}_{v_p},
    \vec{x}_{v_c}, t \right) dt 
\label{eq:BM_InDirect}
\end{aligned}
\end{equation}

where $M_2 (t)$ is the complement of $M_1(t)$, and $M_1(t)$ is given by:

\begin{equation}
\begin{aligned}  
M_1(t)=
\begin{cases}
1, & t \leq t_i \\
0, & t > t_i \\
\end{cases}
\label{eq:BM_Direct_Mask}
\end{aligned}
\end{equation}

We note that $M_1(t)$ and $M_2(t)$ needs to be computed only once per hidden scene with $O(N^3)$ complexity, where $N$ is the side length of reconstruction volume. In Section \ref{sec:Complexity}, we discuss how our algorithm has significantly lower complexity than previous work \cite{marco_virtual_2021}. 

\subsection{Dual Photography and Denoising}
In the phasor field framework, Helmholtz reciprocity is used to treat each SPAD pixel $x_p$ is treated as a virtual illumination source, and each laser position $x_c$ is a point on the virtual aperture that captures the virtual wavefront due to the higher density of the laser grid. Therefore, existing 3D reconstructions for non-confocal NLOS imaging systems with single-pixel ToF detectors are usually \textit{dual images} \cite{sen_dual_2005} that image the hidden scene from the perspective of the laser grid/illumination positions. 
Although fast reconstruction algorithms can be implemented \cite{liu_phasor_2020} on chip \cite{jiang_ring_2022, liao_fpga_2022}, the speed of the imaging pipeline is often restricted by sequential scanning of the laser (Non-Confocal 1 in Fig. \ref{fig:NLOS_Panel}C) which usually involves a mechanical galvanometer. Non-Confocal 2 in Fig. \ref{fig:NLOS_Panel}C shows an equivalent non-confocal acquisition scheme for the \textit{primal} image, where we fix a single laser position and image from the perspective of detector pixels. As demonstrated in Supplement Section \ref{sec:Primal_Dual}, the primal and dual reconstructions are similar since both image the hidden scene from the perspective of the relay wall. However, the second scheme  enables parallel acquisition by leveraging multiple SPAD pixels to collect photons simultaneously, thus improving acquisition efficiency. 

In Supplement Section \ref{sec:SNR}, we conduct an SNR analysis to demonstrate this improvement qualitatively and quantitatively. We show that increasing the number of SPAD pixels with a fixed exposure time is equivalent to extending the exposure time for a single pixel because both methods lead to capturing an increased number of photons. Consequently, using more pixels allows for reduced acquisition time by capturing more photons in parallel.

\begin{figure}[t]
  \centering
   \includegraphics[width=0.9\linewidth]
   {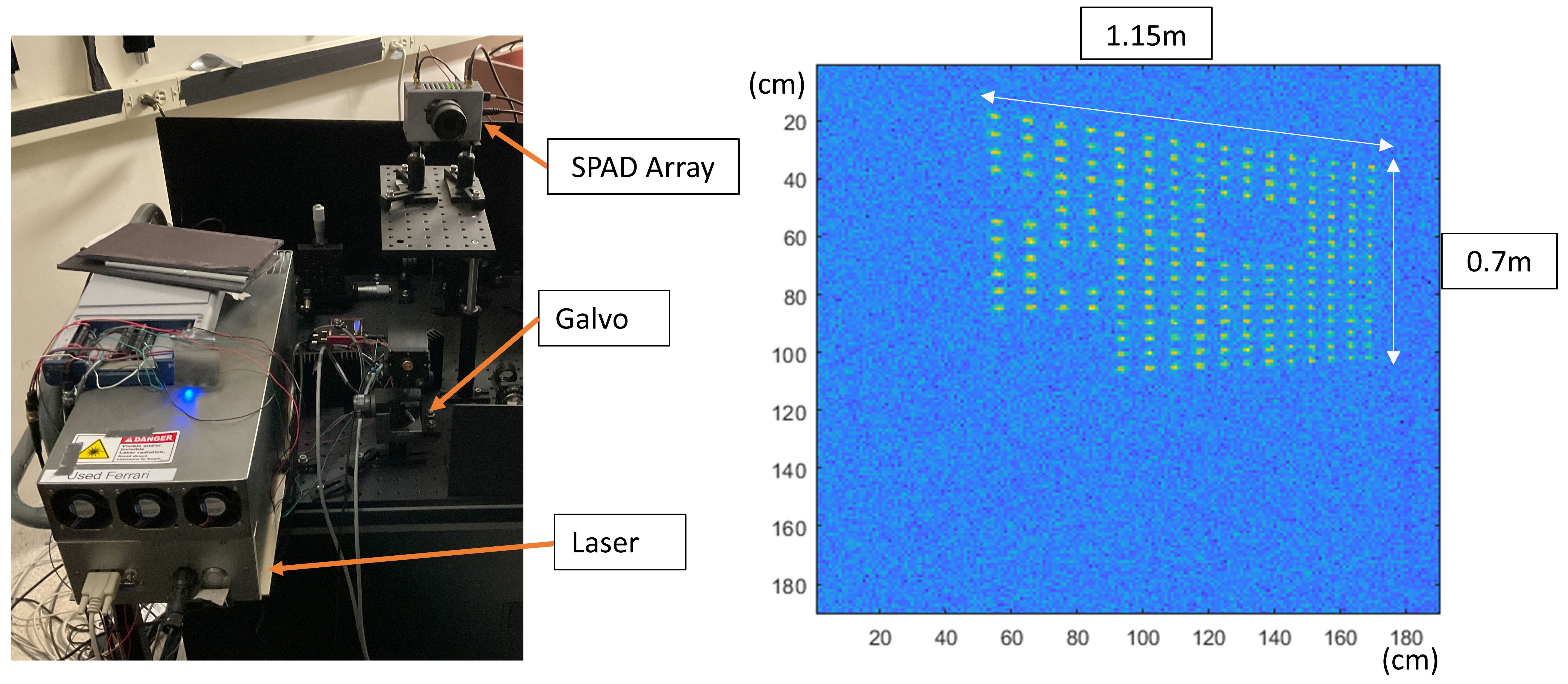}
 %{figures/Final/Hardware_Trimmed.png}
   \caption{Our NLOS imaging system consisting of laser, galvanometer, and SPAD Array. The right figure shows the SPAD pixels focused on the relay surface for large field of view, mimicking a virtual phased array system at the relay wall.}
   \label{fig:LFOV}
\end{figure}

\section{Hardware}
Our NLOS system is an active imaging system (Fig. \ref{fig:NLOS_Panel}B) that consists of the illumination source and a ToF detector. Our illumination source is a \text{PM-1.03-25\textsuperscript{TM}} laser from Polar Laser Laboratories coupled with a frequency doubler that generates 515 nm pulses with a pulse width smaller than 35 ps. The laser operates in burst mode, where the burst repetition rate is 357.14 kHz, and each burst emits 14 pulses with an intra-burst repetition rate of 10 MHz. This configuration yields an effective average repetition rate of 5 MHz across bursts and provides an average output power of 375 mW. To prevent pile-up \cite{becker_advanced_2005}, we use ND filters to adjust the output power for different lens configurations across experiments, resulting in an average power range between 100 mW and 375 mW.

A two-mirror Thorlabs galvanometer (GVS012) is used to scan our relay surface. While the total scan time varies by experiment, the scan area is fixed at a 1.9 m x 1.9 m planar relay wall. For photon detection, we use a customized, gated 16 x 16 SPAD array \cite{riccardo_fast-gated_2022} with a deadtime of less than 100 ns and a temporal resolution characterized by a Full-Width at Half Maximum (FWHM) of approximately 60 ps. This SPAD array is positioned approximately 2.7 m from the relay surface, with different lens configurations employed to span various fields of view across the surface, allowing us to control the resolution and coverage for different experiments.

For SNR experiments outlined in the Supplement Section \ref{sec:SNR}, we use a Canon EF 85 mm f/1.8 USM Lens to focus to a small 7.1 cm x 4.7 cm (Width x Height) area of the relay surface. To perform beamforming and simultaneously image our hidden scene, we need the spacing between the SPAD pixels to be roughly the same as the spacing between laser positions. For this, we use an Edmund Optics 6X Manual Zoom Video lens with the focal length set to 30 mm, which focuses the SPAD array to a 50 cm x 32.5 cm (Width x Height) area on the relay wall. To image the hidden scene under novel illumination and demonstrate dual photography, we use a Fujinon 3 MP Varifocal Lens (3.8-13 mm, 3.4x Zoom) which focuses on a 1.1 m x 0.7 m area on the relay wall (Figure \ref{fig:LFOV}). This figure also shows the "holes" arising from the non-functional SPAD pixels, as well as the distortion in spacing arising from projecting the pixels from one side.

\section{Results}
\begin{figure*}[t]
  \centering
   % \includegraphics[width=0.7\linewidth]{figures/tof_office_iso_216_compare.png}
   \includegraphics[width=0.9\linewidth]
   {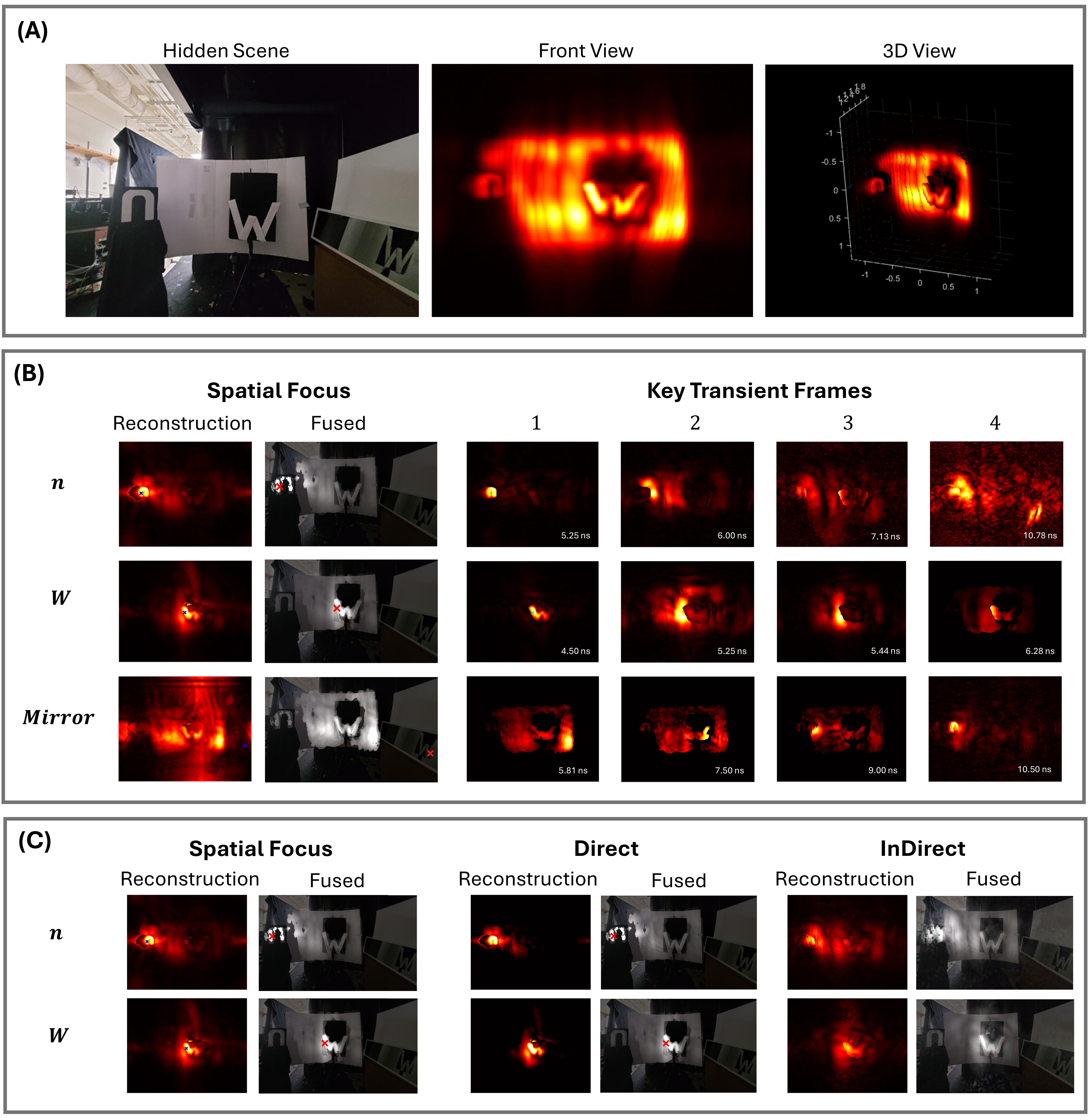}
   % {figures/TMP/TLTM_Panel_Cropped.png}
%   \caption{When we focus the virtual wavefront on the mannequin, then the chair and the shelf do not light up anymore at the correct times. There is some out o}
    \caption{\textbf{(A)} Hidden scene shown on the left. The non-transient, 3D reconstruction generated is shown in the middle (front view) and on the right (3D view). \textbf{(B)} We display 3 columns of TLTM-2 for three locations in the hidden scene from Panel(A). Row 1, 2, and 3 show the impact of focusing the wavefront at different 3D locations corresponding to letters $n$, W, and the mirror in the hidden scene. Columns 1 and 2 mark the focal spot using a blue and red cross respectively, and are generated by adding up all transient frames. Column 2 has the reconstruction overlaid on the ground truth image. Columns 3, 4, 5, 6 correspond to key frames within the video. \textbf{(C)} The TLTM is used to separate direct components (Columns 3,4) and indirect component (Columns 5, 6) of light when focusing the wavefront at locations corresponding to $n$ (Row 1) or W (Row 2). Columns 1 and 2 are generated by adding the direct and indirect components.}
   \label{fig:TLTM_Results}
\end{figure*}

\subsection{TLTM-2 Extraction}
\label{sec:TLTM_Extract}
Our hidden scene consists of the letters $n$, W, a diffuse back wall, and a specular mirror. The scene and the 3D reconstructions using 3rd bounce light is shown in Fig \ref{fig:TLTM_Results}A. Note that mirror does not show up in the reconstruction since it is in the "missing cone" \cite{liu_analysis_2019, royo_virtual_2023} i.e. there are limited number of 3rd bounce light paths that start from the relay wall, interact with the mirror, and end back at the relay wall.

We sequentially focus our virtual illumination to different points $\vec{x}_p^{(2)}$ in the hidden scene, and extract the second order TLTM of the hidden scene, $H^{(2)}\left(\vec{x}_p^{(2)}, \vec{x}_c^{(2)}, t\right)$, using Eq \ref{eq:BM_Video}. In Figure \ref{fig:TLTM_Results}B, we show the effect of focusing our illumination at three different points in the hidden scene. Row 1 shows the effect of focusing our illumination to the letter $n$, lighting up the $n$ (column 3), then hitting the backwall (column 4). The letter $n$ reflects light and illuminates the left side of the letter W (column 5) before illuminating the mirror (column 6). We see an image of the letter $n$ form on the mirror (bottom right of row 1, column 6). We can also image the reciprocal path of focusing on the mirror (Row 3). While we don't see the mirror light up since it is in the missing cone when illuminated from the relay wall, we see the indirect 4th bounce of the illumination from the mirror to the right side of the back wall (column 3), the right side of letter W (column 4), the left side of the back wall (column 5), and finally the letter $n$ (column 6). Row 2 demonstrates the effect of focusing on letter W so W lights up (column 3), then the back wall (columns 4 and 5) and finally the reflection from the back wall lights up the left side of the letter W (column 6). Column 4 in Row 2 is at the same time as Column 3 in Row 4, showcasing that the letter $n$ does not light up when focusing on W. 

\begin{figure*}[t]
  \centering
   % \includegraphics[width=0.7\linewidth]{figures/tof_office_iso_216_compare.png}
   \includegraphics[width=0.85\linewidth]{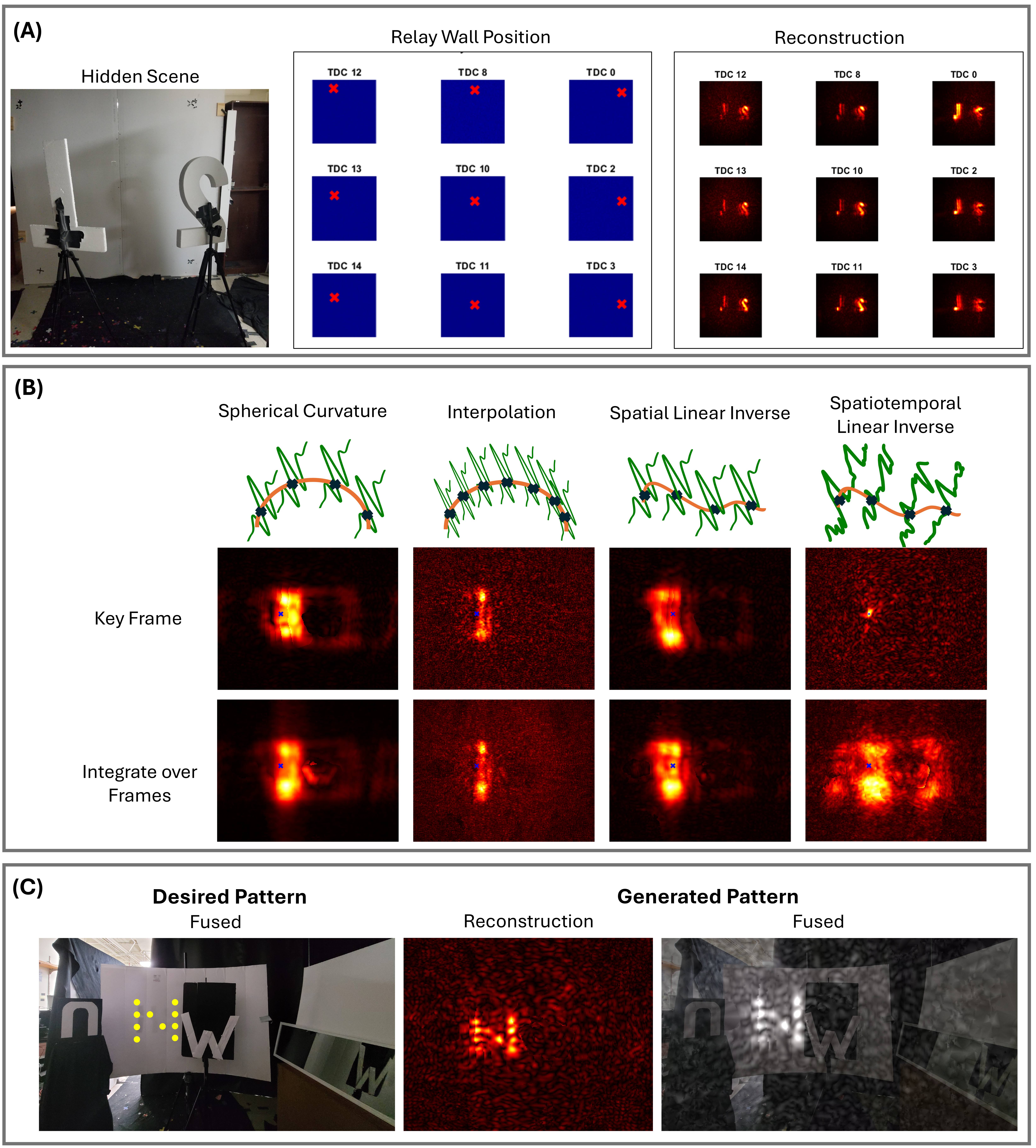}
   %{figures/TMP/Scene_Relighting_Cropped.png}
%   \caption{When we focus the virtual wavefront on the mannequin, then the chair and the shelf do not light up anymore at the correct times. There is some out o}
    \caption{\textbf{(A)} The left image shows the hidden scene while the middle image rough location for the illumination on the relay surface. The image on the right shows the front view of the 3D light field. Each single image is generated by averaging reconstructions over a 4x4 subgroup of neighboring SPAD pixels that share a TDC, and thus each image is labeled using the corresponding TDC index.\textbf{(B)} We demonstrate the impact of focusing a linear array on a point on the back wall, marked by a blue cross. Adding a spherical curvature to our illumination (Column 1) focuses to a vertical line. Interpolation (Column 2) to a finer illumination grid improves the resolution by enabling the use of a shorter phasor field wavelength. Solving an additional linear inverse and optimizing the illumination spatially reduces beamforming artifacts without improving the focus (Column 3). Column 4 shows that we can improve the spatial resolution with which we focus the virtual illumination significantly (Row 1) by sacrificing temporal coherence, which generates random illumination at other frames, as seen in Row 2. \textbf{(C)} By optimizing illumination over both time and space, we can generate incoherent point sources in the hidden scene with high spatial resolution. This allows us to project arbitrary patterns, like the "N" shown in the left image, onto the hidden scene, as displayed in the two images on the right.}
   \label{fig:Scene_Relighting}
\end{figure*}

%A training set is used to optimize the illumination in \textbf{(B)}, Column 3, 4) while an independent test set is used to generate all the results displayed in \textbf{(A)} and \textbf{(B)}

These effects are easier to visualize in a video format, and in supplementary videos SV \ref{vid:BM_FOCUS_SPATIAL} and SV \ref{vid:BM_FOCUS_SPATIAL_Raw}, we showcase the effect of moving the spatial focus to different $\vec{x}_p^{(2)}$ while Videos SV \ref{vid:BM_FOCUS_Video_n}, and SV \ref{vid:BM_FOCUS_Video_W}, and SV demonstrate transient videos when focusing the illumination at key spatial locations.  Additionally, in supplement section \ref{sec:Supp_Results} and supplementary videos SV \ref{vid:Complex_Light_Transport_Water} and SV \ref{vid:Complex_Light_Transport_Milk}, we demonstrate how focusing on the virtual illumination helps reveal shadows in higher-order bounces as well as difference in scattering properties of water and milk, demonstrating that theTLTM-2 preserves complex light transport in the hidden scene.

Our focal spot is elliptical in shape due to the different diameters of the SPAD grid in each dimension, which spans 50 cm x 32.5 cm. When generating a single column of the TLTM-2, we observe residual illumination near the focus point, $\vec{x}_p^{(2)}$, especially in later frames when the virtual illumination should have already passed through. An example of this residual leakage is visible in Row 1, Column 4 of Fig. \ref{fig:TLTM_Results}B, near the letter '$n$'." We believe that this is due to the non-uniformity in the SPAD grid, which arises from the focusing optics and "holes" in the SPAD array (see Fig \ref{fig:LFOV}). To address this non-uniformity, one option is to interpolate across $\vec{x}_p$ to a uniform grid. Alternately, solving an additional linear inverse for side lobe suppression is able to find the optimal spatial curvature that mitigates this residual illumination.

\subsubsection{Improving TLTM-2 Extraction} 
\label{sec:TLTM_Improve}
To improve the quality of the generated TLTM-2 while the spatial resolution is constrained by limited number SPAD pixels, we utilize the methods outlined in Section \ref{sec:BM_Res}. To demonstrate this improvement efficiently, we isolate a 1x16 row of our 16x16 SPAD array and focus to a point on the back wall in the hidden scene, marked by a blue cross in each figure in Fig. \ref{fig:Scene_Relighting}B. The dataset used to generate this figure is captured with an average SPAD pixel spacing of 3 cm, imposing a Nyquist phasor field wavelength of 6 cm. In this case, the linear array is able to focus to a vertical line near the blue cross at the back wall (Fig. \ref{fig:Scene_Relighting}B, Column 1). Grid interpolation reduces pixel spacing and enables the use of a 3 cm wavelength. The narrower vertical line demonstrates that the improvement in spatial resolution, but interpolation on noisy data causes speckle-like artifacts in the reconstruction (Column 2). Although increasing exposure time reduces these artifacts, it is not always feasible.

A different way to improve the quality of TLTM-2 with limited exposure times is to solve an additional linear inverse to reduce side lobes in our virtual illumination. To prevent the method from overfitting to the noise in the data, we split the dataset into two sets - the training set is used to generate optimal projector function coefficients (\ref{eq:Projector_Space}) using Eq. \ref{eq:BM_Pattern_Space}, while the test set is used for generating the all the results displayed in Fig. \ref{fig:Scene_Relighting}B. Row 1 shows the keyframe at the time when the illumination focuses to the blue cross, while Row 2 displays output after summing over all frames. Optimizing the spatial wavefront mainly reduces the out of focus illumination on the back wall and W (Column 3, Fig. \ref{fig:Scene_Relighting}B) in both the keyframe (Column 3, Row 1) and sum over all frames (Column 3, Row 2) when compared to adding a spherical curvature to focus (Column 1). Additionally, this also reduces the residual illumination discussed in the previous section. 

\subsection{TLTM-2 Applications}
\subsubsection{Separating Direct and Indirect Light Transport}
We utilize the time dimension to gate out the direct illumination (\textit{3rd bounce}) from indirect (\textit{higher order bounces}). We focus on n (Row 1) and W (Row 2), and see that the $n$ and $W$ light up in the direct component (columns 3 and 4). The indirect components (columns 5 and 6) display additional bounces. For example, light scatters from the $n$ and hits left side of $W$, and various places for the back wall (Row 1, Columns 5 and 6). When focusing on W, the light propagates to W, then lights up the back wall, before finally lighting up the edge of W again (Row 2, Columns 5 and 6). Since the mirror is in the missing cone, there is no direct component and so it does not make sense to run the algorithm for this scenario. One limitation of this approach is that it assumes ideal focusing for our illumination, so the direct component of light transport leaks into the indirect component when we don't focus to a point. For example, we see the back wall lights up in the indirect components when focusing on both the $n$ and $W$, as this occurs \textit{after} the illumination reaches $n$ and $W$ respectively and because the spatial extent of the focal spot is larger than $n$ and $W$. 

\subsubsection{Scene Relighting}
\paragraph{Novel Illumination} The diversity of information collected is greatly enhanced as different parts of the hidden scene reflect light to different parts of the relay wall. This is equivalent to illuminating the hidden scene from different angles from the relay surface, and then generating a 3D reconstruction for each illumination position. We can either add all the images together \iffalse{to yield 3D lighting and shading for the hidden scene}\fi as shown in Fig \ref{fig:Shading_3D} or we can stack individual 3D reconstructions to build a 3D light field as displayed in Fig. \ref{fig:Scene_Relighting}A, Column 3 where each illumination location is shown in the middle image. Since each SPAD pixel acts as a virtual illumination source, we can demonstrate the impact of illuminating the hidden scene from Figure \ref{fig:Shading_3D} from different parts of the relay wall by focusing the SPAD array on a large area on the relay surface (Fig. \ref{fig:LFOV}, Column 2). Each reconstruction (and illumination) is averaged over 4x4 subgroup of neighboring SPAD pixels to improve SNR. These pixels share a single Time-to-Digital Converter (TDC), and are labeled using their TDC index. We note that the 2 is seen most clearly by the bottom left pixels while the T is seen the most clearly by top right pixels on the relay surface. This is because the opposing tilts of the hidden objects mean that T is in the missing cone \cite{liu_analysis_2019, royo_virtual_2023} of the pixels located on the right-hand side of the relay surface while the 2 is in the missing cones of the pixels on the left. 

\paragraph{Synthesized Illumination} Once we measure the TLTM-2, we can optimize projector functions to create synthetic coherent, point sources in the hidden scene. However, as seen in Fig. \ref{fig:TLTM_Results} (and Video SV \ref{vid:BM_FOCUS_SPATIAL}), the spatial resolution is too low to generate meaningful patterns in the hidden scene. While the spatial resolution can be improved using interpolation (Fig. \ref{fig:Scene_Relighting}B, Column 2) or using more SPAD pixels (Eq \ref{eq:Focusing_Res_SPAD}), another way to do so is to optimize the projector over both time and space to focus to a point. This substantially improves spatial resolution for the keyframe (Fig. \ref{fig:Scene_Relighting}B, Column 4, Row 1), even when compared to the interpolation (Fig. \ref{fig:Scene_Relighting}B, Column 2). However, this technique generates an incoherent point source, sacrificing temporal coherence and introducing out of focus illumination at other frames, which is evident when we sum over frames (Fig. \ref{fig:Scene_Relighting}B, Column 4, Row 2). This substantial improvement is enough to project arbitrary patterns in the hidden scene, where the pattern can be decomposed into incoherent point light sources. To showcase this, we project individual points that form the letter 'N' onto the back wall (Fig. \ref{fig:Scene_Relighting}C, Column 1). We split the dataset into two - the test set is used to generate the projector function using Eq \ref{eq:A_Matrix_Space_Time} $-$ \ref{eq:Projector_Space_Time}, and the test set is used to generate the reconstruction for a single point source using Eq. \ref{eq:PF_RSD_SS_Patterns}. Then, the individual reconstructions are combined using Eq. \ref{eq:Pattern_Normalize}, and the results are displayed in Columns 2 and 3 in Fig. \ref{fig:Scene_Relighting}C. The creases in the back wall impact the projection of the letter "N".

\section{Discussion}
In this work, we have shown that TLTM-2 can be extracted from TLTM-1. However, for a surface 2 m from the relay wall, the spatial resolution of TLTM-2, as described by Eq. \ref{eq:Focusing_Res}
, is an order of magnitude lower than that of TLTM-1 with our current NLOS imaging system. As demonstrated by Eq \ref{eq:Focusing_Res_SPAD}, we can improve the focusing resolution by increasing the number of pixels in the SPAD array. For instance, focusing a 100x100 SPAD array and laser grid across a 2 m relay surface with 2 cm spacing allows for using a 4 cm phasor field wavelength via the nyquist criterion \cite{liu_non-line--sight_2019}. With this wavelength, we can extract TLTM-2 with around 5 cm spatial resolution for a surface 2 m away from the relay surface, which is around 2.5x worse than that of the spatial resolution of TLTM-1. The temporal resolution for TLTM-2 is governed by the width of the phasor field packet, (see \ref{eq:PFPacket_time}), and has a lower bound set by the phasor field wavelength, i.e. 4 $cm$ using the Nyquist criterion. As SPAD arrays with additional pixels become available, both our ability to capture multiple photons simultaneously (see Fig. \ref{fig:snr_quant}) and our spatial resolution with which we extract TLTM-2 will improve linearly. This will enable reconstruction around two corners or increase the resolution with which we reconstruct around the first corner. These additional pixels, coupled with the same fast computational algorithms outlined in this paper, will also enable the extraction of TLTM-3 from TLTM-2 in a feasible manner.

\section{Conclusion}
In this work, we utilize beamforming principles and the phasor field framework to develop fast algorithms that efficiently extract the 7D Transient Light Transport Matrix of the hidden scene $H^{(2)}(\vec{x}_p^{(2)}, \vec{x}_c^{(2)}, t)$ from the Transient Light transport matrix of the relay surface,$H(\vec{x}_p, \vec{x}_c, t)$. Once we compute this TLTM, we can generate the 7D spatiotemporal light transport within the hidden scene. This formalism can be used to build higher order TLTM, $H^{(n)}(\vec{x}_p^{(n)}, \vec{x}_c^{(n)}, t)$, and reconstruct around multiple corners in the future, which will become viable as the number of pixels increase in upcoming SPAD arrays. On the other hand, the increased amount of SPAD pixels comes at the cost of increased data rates, meaning that data processing will become a major bottleneck. Future work will explore how we can compress or reduce our measurement without losing our ability to extract the transient light transport matrix.

\section{Acknowledgments}
This work was supported by the Air Force Office for Scientific Research (FA9550-21-1-0341) and the Defense Advanced Research Projects Agency (DARPA) through the DARPA REVEAL Project HR0011-16-C-0025. PP’s contribution is supported by the US Office of Naval Research under award No. N00014-21-1-2469 and by the US Joint Directed Energy Transition Office (JDETO).

\clearpage
{
    \small
    \bibliographystyle{IEEEtran} 
    \bibliography{main, references_Zotero_T}
}

\onecolumn
\clearpage
\setcounter{page}{1}
\setcounter{section}{0}

{
\newpage
    \onecolumn
    \centering
    \Large
    % \textbf{\thetitle}\\
    \textbf{Transient Light Transport Matrices for Non-Line-of-Sight Imaging} \\
    \vspace{0.5em}Supplementary Material \\
    \vspace{1.0em}
    %< twocolumn
}

\makeatletter
\renewcommand \thesection{S\@arabic\c@section}
\renewcommand\thetable{S\@arabic\c@table}
\renewcommand \thefigure{S\@arabic\c@figure}
\makeatother

In Section 1, we derive camera functions, $\Phi(.)$, that enable 3D and 4D reconstructions for the hidden scene. In section 2, we derive two projector functions: one for reconstructing the hidden scene under novel illumination and another for extracting TLTM-2 from TLTM-1 by using beamforming principles to synthesize a virtual wave that focuses in space-time. Section 3 introduces mathematical tricks to accelerate 4D reconstruction, making the TLTM-2 computation feasible by lowering the computational complexity and preserving the memory complexity. In Section 4, we discuss the computational complexity of computing an additional linear inverse using TLTM-2 to optimize projector functions for generating synthetic coherent and incoherent illumination in the hidden scene. Section 5 examines how increasing the number of SPAD pixels is equivalent to increasing exposure times. Finally, Section 6 presents additional experiments that demonstrate complex light transport is preserved in TLTM-2.

\section{Phasor Field Imaging}
\label{sec:PF_Derivations}

In this section, we derive the camera functions in Eq \ref{eq:PF_Camera} that enable the generation of 3D and 4D reconstructions. We can use the filtered backprojection algorithm \cite{velten_recovering_2012}, to reconstruct each 3D voxel, $\vec{x}_{v}$, as a function of time $t$. This algorithm accounts for the time shifts to each voxel in our hidden scene from each combination of illumination position $\vec{x}_{p}$ and detector position $\vec{x}_{c}$ at the relay surface: 

\begin{equation}
\begin{aligned}
    % \label{eq:I_Supp}
    I_\mathcal{P}\left(\vec{x}_{v}, t\right) & \triangleq \iint \left[ \mathcal{P}(\vec{x}_{p}, t)*H\left(\vec{x}_{p}, \vec{x}_{c}, t \right) \right] \delta\left(t-\frac{|\vec{x}_{p}-\vec{x}_{v}|+\left|\vec{x}_{c}-\vec{x}_{v}\right|}{c} \right) d \vec{x}_{c} d \vec{x}_{p}
    \label{eq:Video_Voxel}
\end{aligned}
\end{equation}

where $\mathcal{P}(\vec{x}_{p}, t)$ is the phasor field projector function (see Fig S1 in \cite{liu_analysis_2019}). This 4D video can also be can be computed efficiently implemented in the \textit{temporal} frequency, $\omega$, domain:

\begin{equation}
\begin{aligned}
I\left(\vec{x}_{v}, t \right) &=\mathcal{F}^{-1}\left\{I_{\mathcal{F}}\left(\vec{x}_{v}, \omega\right)\right\}  \\
& =\iiint \mathcal{P}(\vec{x}_{p}, \omega) H\left(\vec{x}_{p}, \vec{x}_{c}, \omega\right) e^{-j \frac{\omega}{c}\left(\left|\vec{x}_{p}-\vec{x}_{v}\right|+\left|\vec{x}_{c}-\vec{x}_{v}\right|\right)} e^{j \omega t} d\omega d \vec{x}_{c} d \vec{x}_{p} 
\label{eq:Video_Voxel_Freq}
\end{aligned}
\end{equation}

where we have used the time shift property of the Fourier transform to convert a time shift into a frequency dependent phase shift \cite{goodman_introduction_2005}. After reconstruction, each voxel in the reconstructions lights up at $t=0$ since we have shifted the time index to account for the photon time of flight from the illumination to the voxel and back to the sensor. To generate a transient video of the illumination moving through the hidden scene, we can add a voxel-dependent time shift to generate a global time index $t^\prime = t + |\vec{x}_{p}-\vec{x}_{v}|/c$. For this shifted index, $t^\prime=0$ is when our illumination source starts emitting light and light arrives at each voxel at a time index determined by the speed of light. In the Fourier domain, this is equivalent to adding a frequency-dependent phase shift, $e^{-j \phi(\omega)} = e^{j\omega\frac{\left|\vec{x}_{p} -\mathbf{x}_{ \mathrm{v}} \right|}{c}} $:

\begin{equation}
\begin{aligned}
I\left(\vec{x}_{v}, t^\prime \right) &=\mathcal{F}^{-1}\left\{I_{\mathcal{F}}\left(\vec{x}_{v}, \omega\right)  e^{j\omega\frac{\left|\vec{x}_{p} -\mathbf{x}_{ \mathrm{v}} \right|}{c}} \right\}  \\
& =\iiint \mathcal{P}(\vec{x}_{p}, \omega) H\left(\vec{x}_{p}, \vec{x}_{c}, \omega\right) e^{-j \frac{\omega}{c}\left(\left|\vec{x}_{c}-\vec{x}_{v}\right|\right)} e^{j \omega t^\prime} d\omega  d \vec{x}_{c}  d \vec{x}_{p} 
\label{eq:Video_Voxel_Freq_Offset}
\end{aligned}
\end{equation}

where the phase offset no longer depends on $\vec{x}_p$. The non-transient, 3D reconstruction from prior art \cite{liu_non-line--sight_2019, liu_phasor_2020, nam_low-latency_2021} can be rewritten as the intensity of Eq \ref{eq:Video_Voxel} at $t=0$ i.e we get:

\begin{equation}
\begin{aligned}
I\left(\vec{x}_{v}\right) &=  I\left(\vec{x}_{v}, t = 0 \right)  \\ 
&= \iiint \left[ \mathcal{P}(\vec{x}_{p}, \omega) H\left(\vec{x}_{p}, \vec{x}_{c}, \omega\right) \right] e^{j \frac{\omega}{c}\left(\left|\vec{x}_{p}-\vec{x}_{v}\right|+\left|\vec{x}_{c}-\vec{x}_{v}\right|\right)}d \vec{x}_{p} d \vec{x}_{c} d\omega. \\
\label{eq:Voxel_Freq}
\end{aligned}
\end{equation}

where the inverse fourier kernel, $e^{j\omega t}$, simplifies to $1$ since $t=0$.

\section{Projector Functions}
\subsection{Novel Illumination}
\label{sec:PField_Projector_Function}
The phasor field kernel utilized for phasor field imaging \cite{liu_non-line--sight_2019, nam_low-latency_2021} in prior art is a modulated Gaussian in the time domain with some central wavelength of modulation given by $\omega_0 = 2 \pi c/\lambda_C$ and the FWHM of the gaussian given by $\sigma = 5\lambda_C/c$: 

\begin{equation}
\begin{aligned}
\mathcal{P}\left(\vec{x}_p, t\right) =& 
\delta \left(\vec{x} - \vec{x}_p \right) \mathcal{P}(t) \\
=& \delta \left(\vec{x} - \vec{x}_p \right) \frac{1}{\sqrt{2\pi} \sigma} e^{i \omega_0 t}  e^{-\frac{t^2}{2 \sigma^2}}
\label{eq:PFPacket_time}
\end{aligned}
\end{equation}

First, we show that this projector function for is a bandpass filter in the temporal Fourier domain. This filters our signal using a frequency-dependent complex number:

\begin{equation}
\begin{aligned}
\mathcal{P}_{\mathcal{F}}\left(\vec{x}_p, \omega \right) &= \int \mathcal{P}\left(\vec{x}_p, t\right) e^{-i\omega t} dt \\
&=  \delta \left(\vec{x} - \vec{x}_p \right) \frac{1}{\sqrt{2\pi}  \sigma} e^{-\frac{(\omega - \omega_0)^{2} (\sigma)^2}{2}} \\
& = \delta \left(\vec{x} - \vec{x}_p \right) \mathcal{P}(\omega)
\label{eq:PFPacket_Omega}
\end{aligned}
\end{equation}

This projector function illuminates the hidden scene with a gaussian wave \cite{liu_non-line--sight_2019} propagating from $\vec{x} = \vec{x}_p$ in all directions, starting at $t^\prime = 0$. By adding additional illumination positions $\vec{x}_p$ on the relay surface, we can generate 3D and 4D reconstructions under novel illumination using Eq \ref{eq:Video_Voxel_Freq} and Eq \ref{eq:Video_Voxel}.

\subsection{Extracting TLTM-2 using Delay and Sum Beamformer}
\label{sec:BM_Derivations}
To extract TLTM-2, we need to focus our illumination at a specific point $\vec{x}^{(2)}_p$ in the hidden scene while imaging all the other voxels, $\vec{x}^{(2)}_c$. Each temporal Fourier component of Gaussian wave can be modeled as a wavefront of a coherent,  monochromatic, spherical wave emitted at illumination positions $\vec{x}_p$ and detected at positions $\vec{x}_c$ on the relay wall with wavelength $\lambda_c = 2 \pi \omega_0/c$ \cite{liu_non-line--sight_2019, reza_phasor_2019, reza_phasor_2019-1, teichman_phasor_2019, dove2020theory, sultan_towards_2024}. This enables us to model each $\vec{x}_p$ as an omnidirectional element of a virtual phased array system on the relay surface that emits phasor field waves. Using the delay-and-sum beamforming principles, we introduce timeshifts that depend on $\vec{x}_p^{(2)} $ i.e. $\Delta t \left(\vec{x}_p^{(2)} \right)$, that shape the emitted wavefront to steer and focus the virtual illumination on a specific 3D point, $\vec{x}_p^{(2)}$, in the hidden scene. Thus, our new projector function can be written as:

\begin{equation}
\begin{aligned}
\mathcal{P}_{\mathcal{F}}\left(\vec{x}_p, \omega \right) &= \int \mathcal{P}\left(\vec{x}_p, t - \Delta t \left(\vec{x}_p^{(2)} \right) \right) e^{-i\omega t} dt \\
&=  \delta \left(\vec{x} - \vec{x}_p \right) e^{-j \omega \Delta t \left(\vec{x}_p^{(2)} \right)} \frac{1}{\sqrt{2\pi}  \sigma} e^{-\frac{(\omega - \omega_0)^{2} (\sigma)^2}{2}} \\
& = e^{-j \omega \Delta t \left(\vec{x}_p^{(2)} \right)} \mathcal{P}(\omega)
\label{eq:PFPacket_Omega_Supp_BM}
\end{aligned}
\end{equation}

By linking our time delay to the spatial location of our virtual illumination source on the relay wall, we can define appropriate functions that can steer or focus our virtual beam. 

\subsubsection{BeamSteering}
Our virtual illumination position  $x_p$ is a 3D vector with an x-component, $x_{px}$, y-component $x_{py}$, and z-component $x_{pz}$. The mean illumination location, $\bar{x}_p$, can be considered to be the beam origin and a spherical coordinate system consisting of $(r, \Theta, \Phi)$ can be defined to steer the beam to any voxel location given by $\vec{x}_p^{(2)} = \vec{x}_{v}$, along a radial vector $\Vec{r}$. For this coordinate system $r = |\Vec{r}|=|\vec{x}_{v} - \bar{x}_p|$ is the radial distance,  $\Theta$ is the zenith angle between the z-axis and the radial vector $\Vec{r}$, and $\Phi$ is the azimuthal angle between the x-axis and the orthogonal projection of the radial vector, $\Vec{r}$, onto the relay surface given by $z=0$ plane. Assuming our relay wall is planar, we can steer our beam in 3D by introducing time delays that depend only on $x_{px}$ and $x_{py}$. Then, using beamforming principles from ultrasound phased arrays \cite{schmerr_fundamentals_2015}, we can show that the beam can be steered to a voxel $\vec{x}_{v}$ using the following formulas:

\begin{equation}
\begin{aligned} 
\Theta &= \arccos \left( \frac{x_{vz}-\bar{x}_{pz}}{r} \right) \\
\Phi   &= \arctan \left( \frac{x_{vy}-\bar{x}_{py}}{x_{vx}-\bar{x}_{px}}\right) \\
\Delta t (x_p) &= \left( \frac{(x_{px}-\bar{x}_{px}) \sin{\Theta}\cos{\Phi} + (x_{py}-\bar{x}_{py}) \sin{\Theta}\sin{\Phi}}{c} \right)
\end{aligned}
\end{equation} 

\subsubsection{BeamFocusing and BeamSteering}

To steer a focused beam to a voxel $\vec{x}_{v}$, we introduce time delays that impart a spherical curvature to each virtual illumination point $x_p$ and focuses the beam originating from $\bar{x_p}$ at a focal distance $r = |\Vec{r}|=|\vec{x}_{v} - \bar{x}_p|$:

\begin{equation}
\begin{aligned} 
\Delta t (x_p) &= \frac{1}{c}  \sqrt{
\left( r \sin{\Theta}\cos{\Phi}  - (x_{px}-\bar{x}_{px}) \right)^2 
+ (r\sin{\Theta}\sin{\Phi} - (x_{py}-\bar{x}_{py}))^2 + (r\cos(\Phi))^2}
\label{eq:curvature_parabolic}
\end{aligned}
\end{equation}

\section{Efficient Computation for TLTM-2}
In this section, we demonstrate how we can accelerate the computation of Eq \ref{eq:Video_Voxel_Freq_Offset} in the temporal frequency domain. First, we recognize that the shifted time index, $t^\prime$ removes the phase shift dependence of each voxel, $\vec{x}_v$, on the illumination position so we first integrate along $\vec{x}_p$ to generate $\mathcal{P_F} \left( \vec{x}_c, \omega \right)$. Rearranging Eq \ref{eq:Video_Voxel_Freq_Offset}:

\begin{equation}
\begin{aligned} 
I\left(\vec{x}_{v}, t^\prime \right)
& =\iint  \underbrace{\left[\int \mathcal{P}(\vec{x}_{p}, \omega) H\left(\vec{x}_{p}, \vec{x}_{c}, \omega\right) d \vec{x}_{p} \right]}_{\mathcal{P_F} \left(\vec{x}_c, \omega \right)} e^{-j \frac{\omega}{c}\left(\left|\vec{x}_{c}-\vec{x}_{v}\right|\right)}  e^{j \omega t^\prime} d \vec{x}_{c}  d\omega \\
&= \int \underbrace{\left[\int \mathcal{P_F}(\vec{x}_{c}, \omega) e^{-j \frac{\omega}{c}\left(\left|\vec{x}_{c}-\vec{x}_{v}\right|\right)}  d \vec{x}_{c}  \right]}_{R_{\vec{x}_{v}} \left( \mathcal{P_F} \left(\vec{x}_c, \omega\right)\right)} e^{j \omega t^\prime} d\omega 
\end{aligned}
\end{equation}

Provided all the detection points, $x_c$, lie on a regular grid on a planar surface, then the the inner integral can be written as a convolution: 

\begin{equation}
\begin{aligned}
        R_{\vec{x}_{\mathrm{v}}}  \left(\mathcal{P}_{\mathcal{F}}\left({\vec{x}}_{\mathrm{c}}, \omega\right)\right) &=  R_{\vec{x}_{\mathrm{v}}}  \left(\mathcal{P}_{\mathcal{F}}\left(
        x_c, y_c, z_v, \omega \right) \right)
       \\
       \\
       &=   \mathcal{P}_{\mathcal{F}}\left(
        x_c, y_c, 0, \omega \right) \underset{\text{2D}}{*} G\left(x_c, y_c, z_v, \omega\right) \\  
\end{aligned}
\label{eq:FBP_PF_Omega_RSD_CONV}
\end{equation}

where $ G\left(x_c, y_c, z_v, \omega \right) = {e^{i\frac{\omega}{c} \sqrt{{x_c}^2 + y_c^2 + z_v^2}}}/{\sqrt{x_c^2 + y_c^2 + z_v^2}} $. The reconstruction for all $\vec{x}_{v}$ at a fixed depth $z_v$ and frequency $\omega$ can plane can be computed simultaneously by using a single 2D FFT which significantly reduces the computational complexity \cite{liu_phasor_2020,  gu_fast_2023}. These mathematical tricks are valid as long as $\vec{x}_c$ is sampled using a regular grid and the projector function, $\mathcal{P}\left(\vec{x}_p, t\right) $, contains no implicit dependence on $\vec{x}_c$. Since this is true for all projector functions discussed in this paper, columns of TLTM-2 and 4D reconstructions under novel illumination can be generated using this fast reconstruction scheme. 

In the next two sub sections, we examine the computational complexity of this algorithm. For this analysis, let $L\times L$ denote the number of laser positions, $M\times M$ denote the number of spad pixels focused on the wall, $T$ denote the number of time bins, and $F$ the number of frequency components.

\subsection{Computational Complexity}
\label{sec:Complexity}
 The computational complexity of the RSD algorithm when reconstructing a volumetric cube of side length $N$ for a single SPAD pixel is $O(N^3 \log{N})$. For $M^2$ SPAD pixels, we run this algorithm $M^2$ times and aggregate the results to yield a complexity of $O(M^2 N^3 \log{N})$. This process can be parallelized to significantly reduce runtime \cite{nam_low-latency_2021}. 
 
To generate a transient video for $k$ total frames, we first compute the inner integral in Eq \ref{eq:Video_Voxel_Freq_Offset} with respect to $\vec{x}_p$. Adding the transient measurements together maintains the memory complexity. Generating each frame has the same reconstruction complexity as the RSD algorithm for a single SPAD, leading to an overall complexity of $O(k N^3 \log{N})$. This is the complexity for generating a single column of the TLTM i.e. for a single spatial focus. To generate the full TLTM, we need to repeat the previous computation for each spatial location in the hidden volume, yielding an overall complexity of $O(SkN^3log(N))$, where $S$ is the number of spatial focus locations. 

The complexity for a single spatial focus, $O(k N^3 \log{N})$, is orders of magnitude faster than previous work \cite{marco_virtual_2021}, where the corresponding complexity is given by $O(N^3L^2M^2)$ and can be approximated as $O(N^7)$ when $N \approx L\approx M$. Notably, the output of prior work corresponds to a column of the virtual light transport matrix, isolating only 4th-bounce light transport without transient temporal information.

\subsection{Memory Complexity}
For sequential reconstruction, the memory complexity remains the same as that of previous algorithms reported for single pixel SPAD \cite{liu_phasor_2020}. For steady-state reconstruction, we load each histogram of dimension $L^2T$ into memory and compute the reconstruction for an $N^3$ volume, which yields a complexity of $O(L^2T + N^3)$. This can be reduced by directly using the temporal Fourier transform of the histogram, yielding $O(L^2 F + N^3)$, where $F$ is typically an order of magnitude smaller than $T$.

For transient video reconstruction with $k$ frames, the memory requirements remain similar to steady-state reconstruction. After applying SPAD-dependent temporal shifts to focus the wavefront, transient histograms from different SPADs are aggregated together and used to generate a 3D volume for each frame. To avoid excessive memory usage, only a 2D surface (the location of maximum intensity along depth) is saved for each frame, resulting in a memory complexity of $O(L^2 F + kN^2)$, thus avoiding the need to store a full 4D volume.

GPU-based reconstruction further accelerates processing but requires additional memory. For example, our GPU implementation generates a single column of TLTM-2 (one spatial focus) in 1-2 minutes for 141 video frames, compared to approximately 30 minutes for MATLAB-based sequential reconstruction for a dataset that is collected with 216 SPAD pixels.

\section{Complexity for Optimizing Projector Functions}
\label{sec:Complexity_Projector}
Let $A$ refer to some $n \times m$ matrix. When $A$ is inverted using ridge regression, the total complexity is $O(n \cdot m^2 + m^3)$ where the bottlenecks in computation arise from computing ($n \cdot m^2$) and inverting ($m^3$) the Gram matrix ($A^TA$). Let $N$ denote the side length of the reconstruction volume, $M$ x $M$ denote the number of SPAD pixels, and $F$ the number of frequency components we retain after applying the phasor field filter in Eq \ref{eq:PFPacket_Omega}. Suppose, we aim to project a pattern that is decomposed into $P$ points arriving at $S$ unique times, where $S \leq P$. Then, both $A$ and $A_{*}$ need to be recomputed for $S$ unique points, meaning the complexity of ridge regression in Eq. $\ref{eq:A_Matrix_Space}$ and Eq. $\ref{eq:A_Matrix_Space_Time}$ also scales by $S$. When constructing the matrices $A_{*}$ and $A$, we apply a max filter along depth to isolate $N^2$ voxels corresponding to surfaces in the hidden scene, onto which we can project the pattern. This also reduces computation by removing voxels with zero or small intensities.

The matrix $A$ in Eq. \ref{eq:A_Matrix_Space} optimizes the projector function spatially across SPAD pixels and has size $N^2 \times M^2$, making the computational complexity for projecting $S$ points to be $O(S(N^2 M^4 + M^6))$. Similarly, matrix $A_{*}$ in Eq. \ref{eq:A_Matrix_Space_Time} optimizes the projector function over both space and temporal frequencies so the size of $A_{*}$ is $N^2 \times M^2F$. The overall computational complexity of projecting $S$ unique points onto surfaces in the hidden scene is therefore $O(S(N^2 M^4 F^2 + M^6F^2))$.

We note that both complexities scales in polynomial time with regards to $M^2$, the number of SPAD pixels, while the complexity for direct focusing using beamforming, given by $O(SkN^3\log N)$, does not.  Since beamforming resolution also improves with the number of pixels, we anticipate that these optimization procedures will become unnecessary as SPAD array pixel counts increase.
 
\begin{figure*}[!t]
\centering
\includegraphics[width=\textwidth]{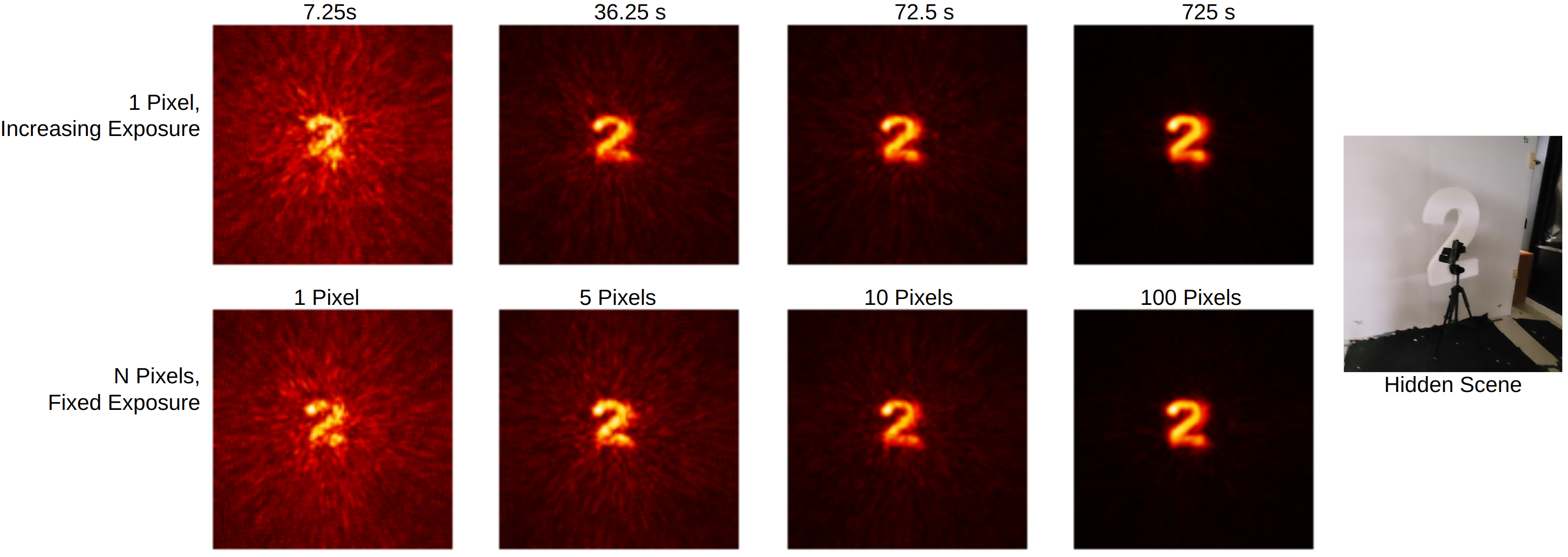}
\caption{Qualitative comparison of SNR. \textbf{Left, Top row:} Single-pixel reconstructions at progressively longer exposure times. \textbf{Left, Bottom row:} Reconstructions with a fixed 7.25s exposure time using an increasing number of SPAD pixels. \textbf{Right:} Hidden object}
\label{fig:snr_qual}
\end{figure*}

.\begin{figure*}[!t]
\centering
\includegraphics[width=0.9\textwidth]{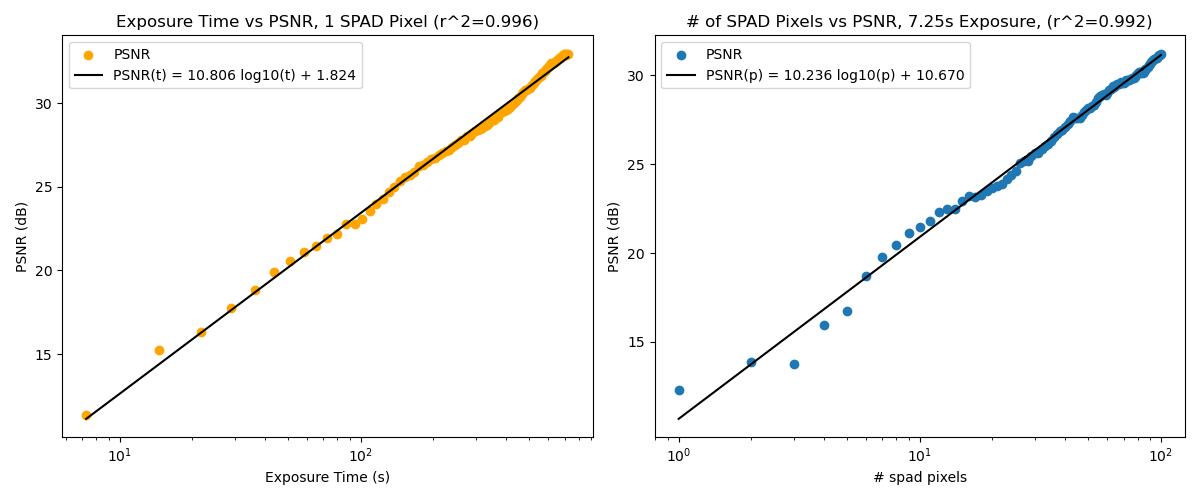}
\caption{Quantitative comparison of SNR, showcasing that the growth factor for PSNR is similar when increasing the exposure time (10.806) or the number of SPAD pixels (10.236).}
\label{fig:snr_quant}
\end{figure*}

\section{SNR}
\label{sec:SNR}

A practical benefit of using a multi-pixel SPAD array is the ability to capture significantly more photons per unit of time than when using a single-pixel photon detector. While this may seem intuitively obvious, we believe the consequences of this benefit are foundational to advancing NLOS research and furthering the development of practical NLOS applications. Therefore, this fact warrants closer inspection. 

A researcher accustomed to imaging with almost any transient camera method is familiar with the unavoidable trade-off: exposure time versus image quality. SNR is largely a function of the number of photons collected. Short exposure times collect fewer photons, therefore have higher Poisson noise, and consequently, lower SNR. Longer exposure times ameliorate the low-SNR problem because Poisson noise grows at a rate proportional to the square root of the number of photons detected. However, long exposure times are not a universal panacea because many compelling practical applications of NLOS imaging either involve dynamic scenes or require near real-time analysis, e.g. \cite{nam_low-latency_2021}.

The expectation is that for a given single-pixel exposure time, $t_1$, an equivalent SNR can be achieved by instead using $n$ pixels with an exposure time of $t_n = t_1/n$. We show for the first time experimentally that this expectation does indeed hold true. To achieve this trade-off, it is necessary to project the SPAD array onto a small area on the relay wall, ideally having dimensions of the same order as the phasor-field wavelength. By doing so, all pixels observe nearly the same perspective of the hidden scene and can be easily combined without introducing any additional blurring in the combined reconstruction. In our experimental data, all SPAD pixels were projected onto a rectangular area of 7.1 $\times$ 4.7 cm.

In our experiment, we constructed a simple scene containing a Lambertian number `2' cut from expanded polystyrene foam. We then captured 125 exposures of 7.25s each using all available SPAD pixels on our array. Analysis was performed in the following manner: First, we selected a single pixel at random from the array. Using only photons collected by this pixel, we performed 100 phasor-field reconstructions having exposure times $7.25 \times n$ seconds for $n = 1\ldots100$. Second, we selected a single 7.25-second exposure, and performed 100 phasor-field reconstructions, using $n$ arbitrarily selected pixels for $n = 1\ldots100$. As expected, we observed a linear increase in total detected photon count with both the increasing exposure and increasing pixels analyses. A sampling of the reconstructions for $n = \{1, 5, 10, 100\}$ are shown in figure \ref{fig:snr_qual}. 

For quantitative analysis, we used the widely accepted Peak Signal-to-Noise Ratio (PSNR) calculation as a metric to assess improvement in image quality with the addition of more photons, whether by increased exposure or increased number of pixels. The PSNR formula requires calculating the mean square error with respect to a noise-free image. For the noise-free image in our calculations, we used a reconstruction from 216 pixels of a 906.25 second exposure. The PSNRs of each of the 200 reconstructions were calculated and are shown in figure \ref{fig:snr_quant}. A least-squares, best-fit exponential curve overlays each data set. Note that both methods of adding additional photons (longer exposure vs additional pixels) have nearly the same growth rate for the reconstructions' PSNR (0.186 vs 0.184). From this, we can conclude that when a tight spacing of SPAD pixels on the relay wall is used, capturing with multiple SPAD pixels is experimentally equivalent to capturing over a longer exposure time.

\subsection{Primal-Dual Equivalence}
\label{sec:Primal_Dual}
The photon time of flight is identical whether a photon travels from the illumination to the detection or from the detection to the illumination. This means that we should generate the same reconstruction for a given hidden scene if we sample the illumination and detection grid at the same locations on the relay surface, showcasing the equivalence of primal-dual images.

Fast reconstruction algorithms utilize Fourier transforms to speed up reconstruction but require sampling the relay wall using a uniform grid. Although this is feasible for the laser grid where we use a galvo to sequential scanning each grid, focusing all the SPAD pixels simultaneously to a uniform grid is difficult since the focusing optics and the presence of hot pixels on the sensor lead to non-uniformity and distortions seen in Figure \ref{fig:LFOV}. Therefore, we use a slow phasor field back projection algorithm when reconstructing the primal image using the SPAD grid to demonstrate the equivalence between the primal and dual images. 

\begin{figure}[h]
\centering
\includegraphics[width=.5\textwidth]
{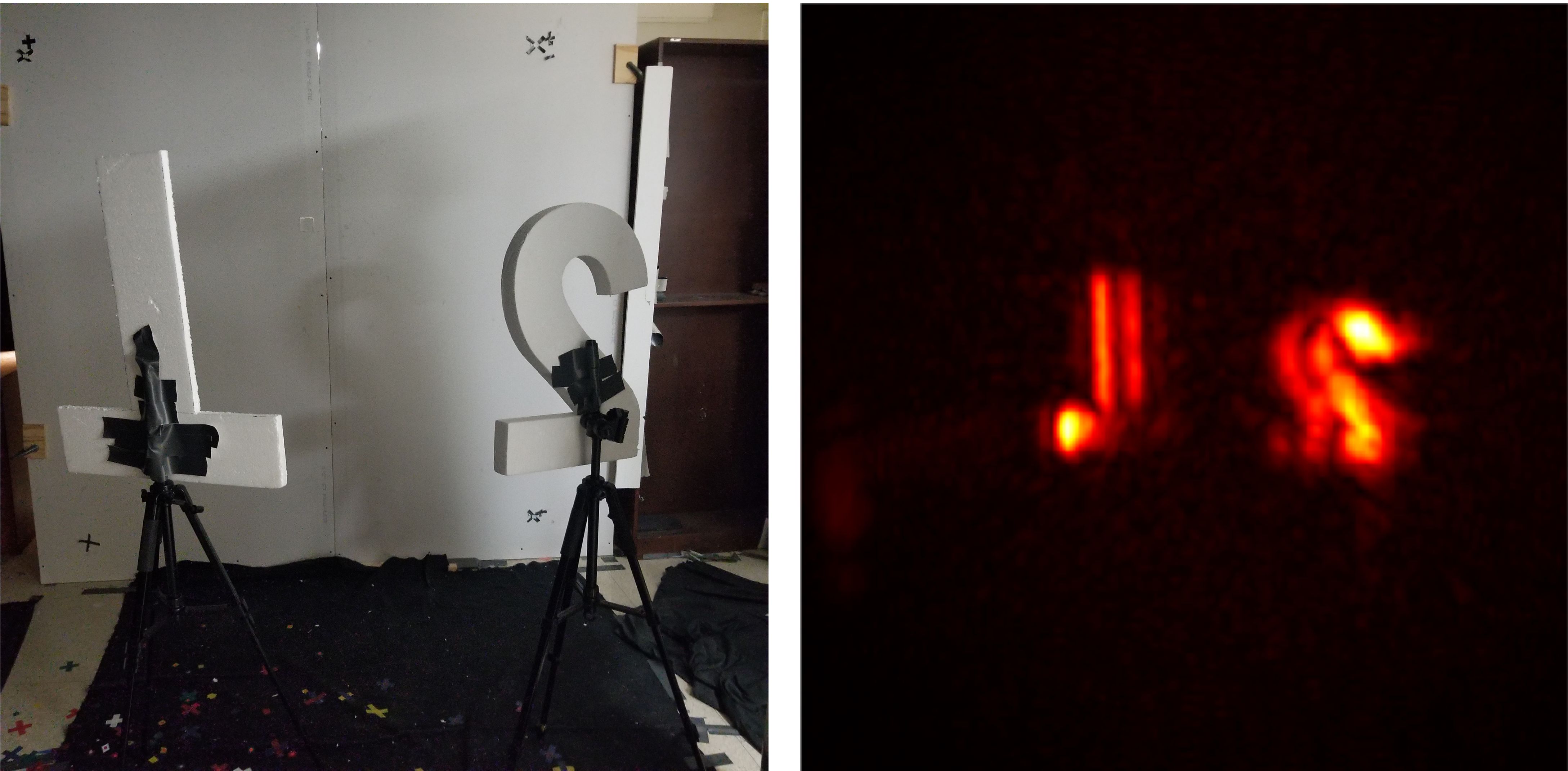}
\caption{Hidden scene shown on the left, and reconstruction from all SPAD pixels for the large field of view camera grid is shown on the right. Adding up all the contributions yields 3D shading and lighting effects for the hidden scene.}\label{fig:Shading_3D}
\bigbreak
\includegraphics[width=.5\textwidth]{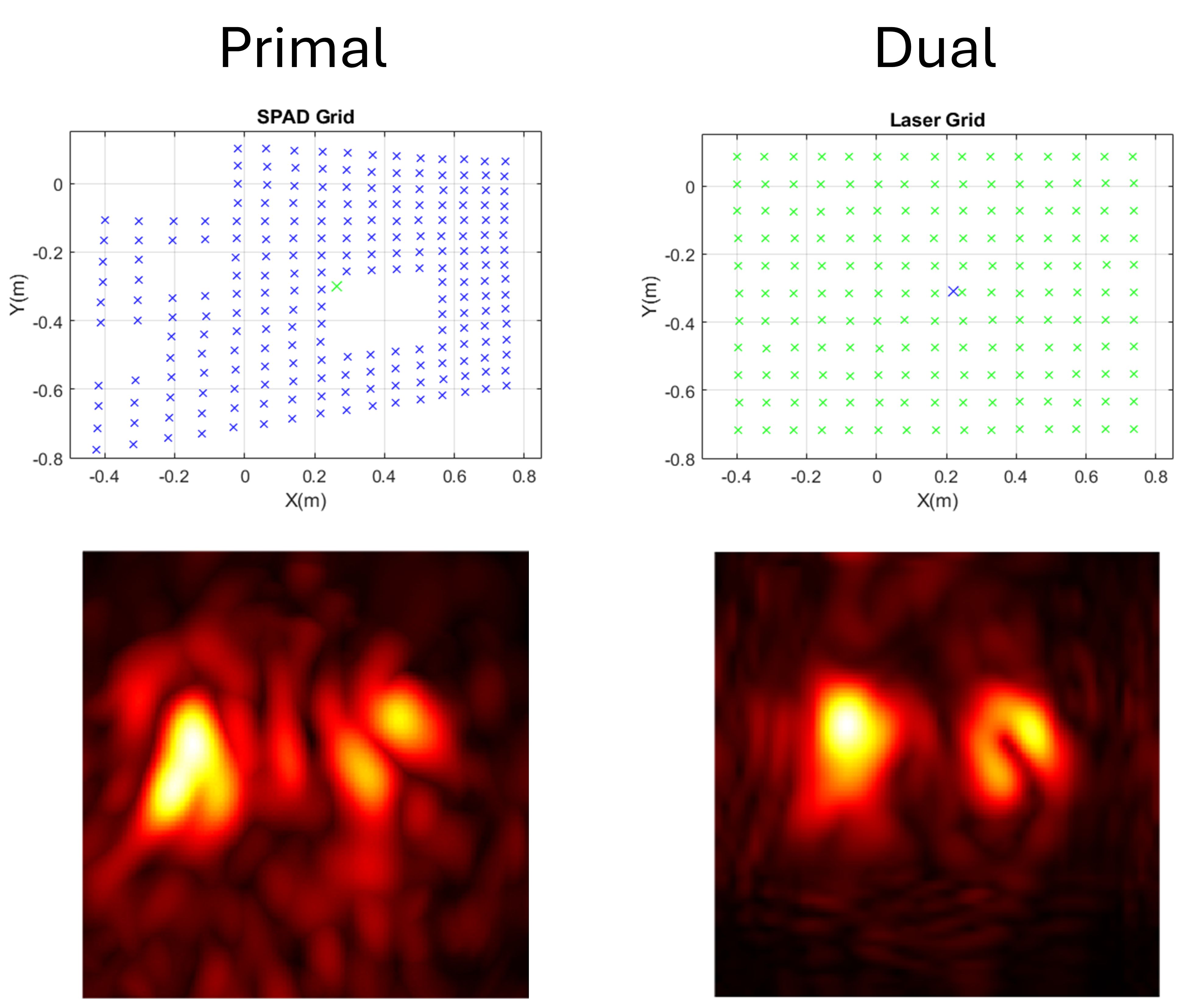}
\caption{Qualitative comparison of primal (column 1) and dual (column 2) images, where the dataset captured using the dual grid takes approximately 250x longer to acquire than that from the primal grid. Row 1 displays the two grids used, with camera positions marked in blue and laser positions marked in green. Row 2 displays the corresponding reconstruction.}\label{fig:primal_dual}
\end{figure}

Figure \ref{fig:Shading_3D} shows the hidden scene (left) and the reconstruction (right) after adding up the reconstructions from all the SPAD pixels for the full laser grid. Next, we reconstruct the primal (left) and dual (right) images for a single laser and camera position respectively, and display them in Fig \ref{fig:primal_dual}. To get the SPAD and laser grids to match (Row 1, Figure \ref{fig:primal_dual}), the laser grid is downsampled leading to a loss of reconstruction resolution. Additionally, the laser positions outside the corners of the SPAD grid are discarded (Figure \ref{fig:LFOV}), which is why the bottom of the 2 and the T is missing in the reconstruction. The 2 in the laser grid reconstruction looks cleaner, but that may be due to the prescence of hot pixels within the SPAD grid.

\section{Complex Light Transport}
\label{sec:Supp_Results}
In this section, we demonstrate that the rich light transport for the hidden scene is preserved in TLTM-2. First, we show how focusing on the virtual illumination helps reveal shadows in higher-order bounces. Second, we demonstrate the difference in scattering properties of water and milk when these are placed in the hidden scene. 

\subsection{Complex Light Transport: 4th Bounce Shadow}
\begin{figure*}[!htp]
  \centering
    \includegraphics[width=0.7\linewidth]{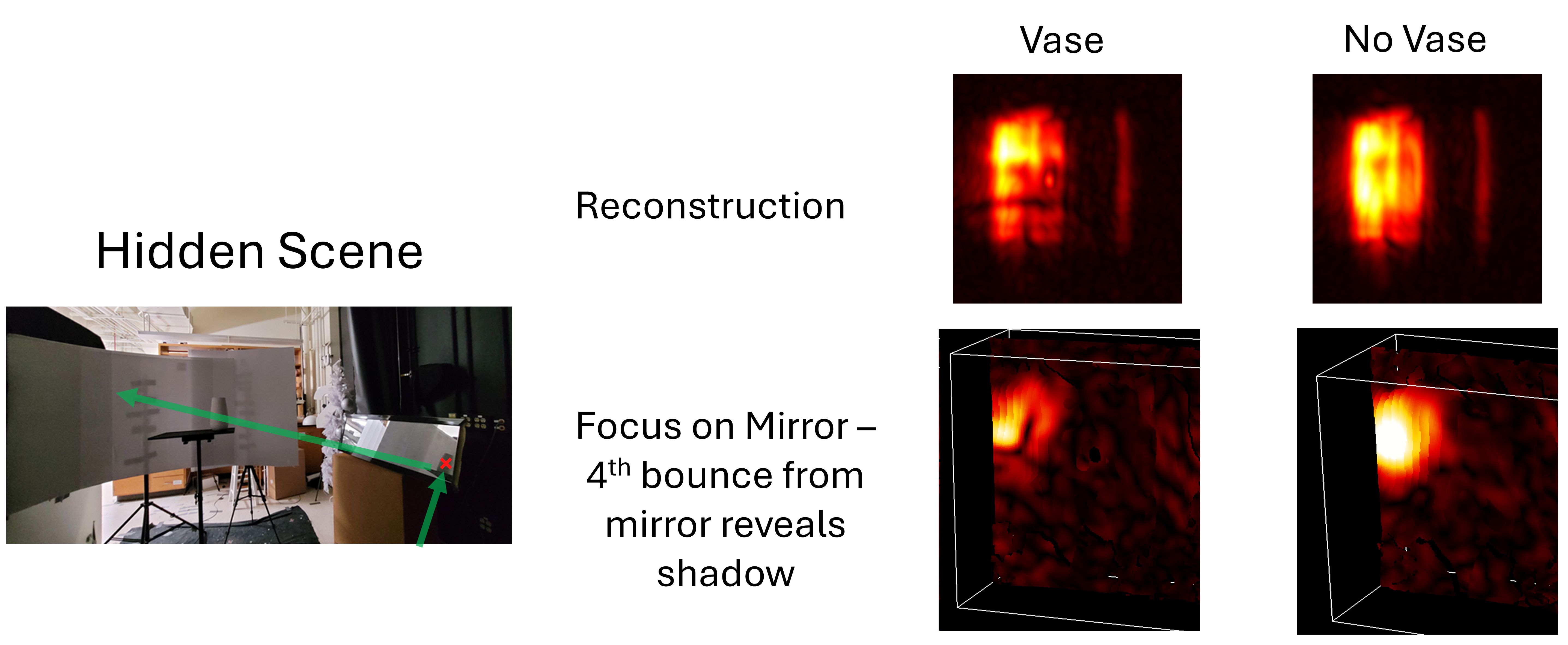}
    \caption{Hidden scene shows on the left, consisting of a vase, back wall, and a mirror. When we focus our virtual illumination at the red cross on the mirror, we see the backwall light up due to the 4th bounce of light from light hitting the mirror and then the back wall (Row 2). The green arrows outline the light path in the left image. We also see a shadow of the vase blocking some of this light (Column 2, Row 2). When we remove the vase, we see that this shadow disappears.}
   \label{fig:BM_4thBounce_Shadow}
\end{figure*}

In Figure \ref{fig:BM_4thBounce_Shadow}, we showcase the light transport effects within the indirect component of the TLTM. Here, we place a vase in front of a back wall, along with a mirror on the side. Focusing our virtual illumination on the mirror reveals a 4th bounce of light hitting the mirror followed by hitting the back wall (Row 2). When the vase is present, a shadow corresponding to the indirect illumination hitting the vase (and the vase stand) can be seen on the back wall (Row 2, Column 1). We verify this by removing the vase and the vase stand, leading to the disappearance of the shadow  (Row 2, Column 2)

\subsection{Subsurface Scattering}
\begin{figure*}[!htp]
  \centering
    \includegraphics[width=0.7\linewidth]{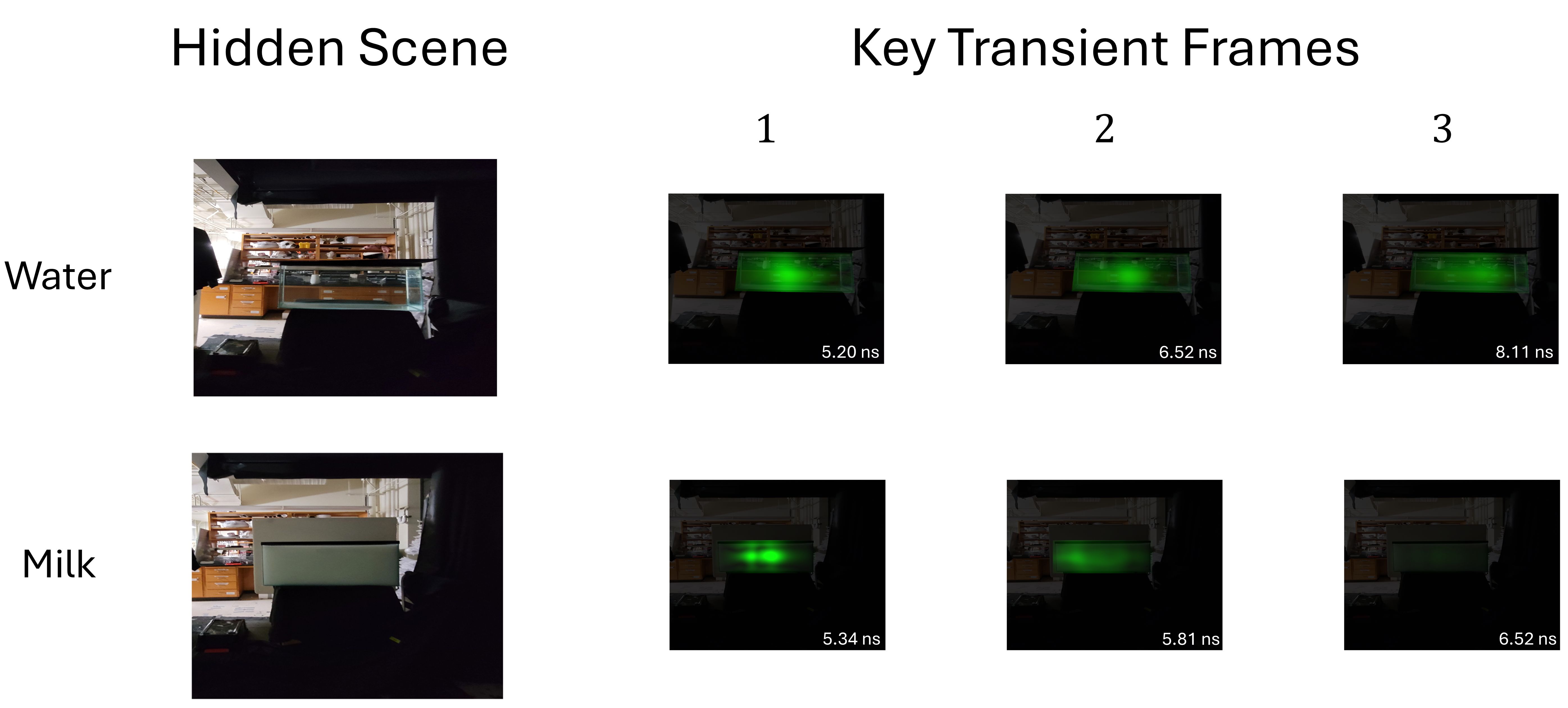}
    \caption{Hidden scene shown on the left. Still images from a spatial-temporal video showing light propagating through the glass tank containing water (Row 1) and milk (Row 2). Top row: Illuminate the tank with a plane wave, lighting up the tank (column 2). Some light propagates through the clear tank and illuminates the back wall (column 3). Finally, we see the reflection from the back wall illuminating the front wall (column 4).  Bottom row: The light coming back is from the milky solution within the tank (column 2), with the illumination persisting within the tank due to subsurface scattering (column 3). No light is reflected back from the back wall of the tank due to the opaque milky mixture (column 4). See supplemental materials for video files of these reconstructions.
    }
   \label{fig:BM_SubSurface}
\end{figure*}

We place an aquarium in the hidden scene and extract a column of TLTM-2 corresponding to light propagating through the tank. \ref{fig:BM_SubSurface} shows key temporal frames of the full column/video. Row 1 shows how the illumination interacts with water, showcasing caustic direct reflections from the front wall of the tank (column 2), the back wall of the tank (column 3), as well as the indirect caustic reflection from the reflection of the back wall hitting the front wall  (column 4). We then add milk, and demonstrate how the illumination interacts differently (Row 2) with the milk-water mixture. We encounter diffuse scattering from within the tank (0.14 ns) after the reflection from glass front (column 2),, and observe subsurface scattering from the milky solution (column 3). Finally, we don't see any reflection from the back wall of the tank at the corresponding frame (column 3). The full videos for water (SV \ref{vid:Complex_Light_Transport_Water}) and milk (SV \ref{vid:Complex_Light_Transport_Milk}) are attached in supplemental materials. 

\end{document}